\documentclass[journal]{IEEEtran}
\usepackage{graphicx}
 \usepackage{float}
 \usepackage{algorithm}
\usepackage{algorithmic}
 \usepackage{subfigure}
\usepackage{amsmath,amssymb,amsfonts}

 \usepackage{url}
 \usepackage{amsmath,bm}
 \usepackage{cite}
 \usepackage{url}
\usepackage{multirow}

\ifCLASSINFOpdf

\else

\fi

\begin{document}

\title{Distributed Online Anomaly Detection for Virtualized Network Slicing Environment}

\author{Weili~Wang,
        Chengchao~Liang,
        Qianbin~Chen,~\IEEEmembership{Senior Member,~IEEE},
        Lun~Tang,
       Halim~Yanikomeroglu,~\IEEEmembership{Fellow,~IEEE}
\thanks{W. Wang, C. Liang, Q. Chen, and L. Tang are with the School of Communication and Information Engineering and the Key Laboratory of Mobile
Communication,
Chongqing University of Posts and Telecommunications, Chongqing
400065, China (e-mail: 1961797154@qq.com; liangcc@cqupt.edu.cn; cqb@cqupt.edu.cn;
tangl@cqupt.edu.cn).}
\thanks{H. Yanikomeroglu is with the Department of Systems and
Computer Engineering, Carleton University, Ottawa, ON, Canada (e-mail:
halim@sce.carleton.ca).}}

\maketitle

\begin{abstract}
As the network slicing is one of the critical enablers in communication networks, one anomalous physical node (PN) or physical link (PL) in substrate networks that carries multiple virtual network elements can cause significant performance degradation of multiple network slices. To recover the substrate networks from anomaly within a short time, rapid and accurate identification of whether or not the anomaly exists in PNs and PLs is vital. Online anomaly detection methods that can analyze system data in real-time are preferred. Besides, as virtual nodes and links mapped to PNs and PLs are scattered in multiple slices, the distributed detection modes are required to adapt to the virtualized environment. According to those requirements, in this paper, we first propose a distributed online PN anomaly detection algorithm based on a decentralized one-class support vector machine (OCSVM), which is realized through analyzing real-time measurements of virtual nodes mapped to PNs in a distributed manner. Specifically, to decouple the OCSVM objective function, we transform the original problem to a group of decentralized quadratic programming problems by introducing the consensus constraints. The alternating direction method of multipliers is adopted to achieve the solution for the distributed online PN anomaly detection. Next, by utilizing the correlation of measurements between neighbor virtual nodes, another distributed online PL anomaly detection algorithm based on the canonical correlation analysis is proposed. The network only needs to store covariance matrices and mean vectors of current data to calculate the canonical correlation vectors for real-time PL anomaly analysis. The simulation results on both synthetic and real-world network datasets show the effectiveness and robustness of the proposed distributed online anomaly detection algorithms.
\end{abstract}


\IEEEpeerreviewmaketitle

\section{Introduction}

\IEEEPARstart{N}{etwork} slicing has been regarded as an important technology to satisfy diverse requirements for emerging service types \cite{ordonez2017network}, \cite{elayoubi20195g}. Network function virtualization (NFV) facilitates the deployment of network slices through unleashing network functions from the dedicated hardware and implementing them on general-purpose servers \cite{oi2015method}. By NFV technique, network slicing can customize virtual network functions based on demands to generate service paths, i.e., customized service function chains (SFCs). Generally, multiple network slices can be instantiated in a shared substrate network to support diversified applications \cite{yousaf2017nfv}, \cite{yang2018stochastic}.

A substrate network consists of multiple physical nodes (PNs) and physical links (PLs), and a virtualized network slice consists of multiple virtual nodes (VNs) and virtual links (VLs). Due to the many-to-one embedding relationship between VNs (VLs) and PNs (PLs), network slices are prone to unexpected and silent forms of failures caused by anomalies arising in the shared substrate network \cite{miyazawa2015vnmf}. As the performance of multiple network slices hinges on the normal running of the shared substrate network, the accurate and rapid anomaly detection for PNs and PLs in the substrate network is the prerequisite for ensuring the service quality of network slices.

Anomaly detection refers to the problem of finding patterns in data that do not conform to the expected behaviors \cite{chandola2009anomaly}. Existing efforts \cite{Callegari2011A,Kun2017fast,9256316,9064715,jiang2013anomaly,schuster2015potentials,9173800,chen2017fault} on anomaly detection usually model all data of a time slot as a long vector and consider it as a single sample in implementing data analysis. Based on this idea, in virtualized network slicing environment, we can obtain the traning dataset for PN anomaly detection through modeling all measurements of VNs mapped to a PN within a period as a series of vectors. Then, normal profiles can be trained to differentiate anomalies arising in PNs. Furthermore, the VN measurements can also be utilized to implement PL anomaly detection. Since the flow passes through each VN of an SFC in sequence, the measurements between neighbor VNs are naturally related \cite{cotroneo2017fault}. Therefore, we can implement PL anomaly detection by analyzing the correlation of measurements between neighbor VNs, which are mapped to both ends of the PL. Through modeling all measurements of VNs mapped to both ends of a PL as two sets of vectors, we can obtain the training dataset for the PL anomaly detection. Note that the key difference between PL and PN anomaly detection is that the anomaly detection for PLs is required to find anomalous correlation patterns between two vectors.

However, since VNs are distributed in multiple slices, if we collect all relevant VN data of a time slot to form vectors and analyze them in a centralized mode, the data privacy of different slices will be compromised. Besides, additional communication and storage cost can be introduced by collecting all data to a central manager. Therefore, distributed detection modes\cite{o2016distributed}, \cite{chen2019distributed}, which can detect anomalies in PNs and PLs by analyzing distributed VN datasets in respective managers, are more effective for virtualized network slicing scenarios.

To recover the substrate networks from anomaly within a short time, the identification of whether or not the anomaly exists in PNs and PLs should be rapid and accurate. However, classical anomaly detection methods are usually based on the batch formulation, which requires storing all historical data within a period and trains the normal profiles through offline learning. The process will not only introduce high storage and computation cost but prevent timely detection of anomalies \cite{li2019online}, \cite{xie2018line}. Besides, in anomaly detection, the training datasets are assumed to contain only one-class samples, i.e., normal samples. If the number of anomalous data contained in training samples increases, the performance of the detection model trained by the offline learning can be greatly degraded \cite{zhang2018boosting}. Therefore, online anomaly detection methods that can analyze system data in real-time are preferred.

Based on the above analysis, we propose a distributed online PN anomaly detection algorithm based on a decentralized one-class support vector machine (OCSVM) and a distributed online PL anomaly detection based on the canonical correlation analysis (CCA). By the distributed online mode, the real-time measurements of VNs can be analyzed in respective managers distributedly to detect anomalies in PNs and PLs. To the best of our knowledge, this is the first work which proposes the distributed online anomaly detection algorithms for the virtualized network slicing environment. Our main contributions are summarized as follows:

Firstly, based on a decentralized OCSVM, we propose a distributed online PN anomaly detection algorithm for the virtualized network slicing environment. Specifically, to decouple the OCSVM objective function, we transform the original OCSVM objective function to decentralized quadratic programming problems by introducing the consensus constraints. To realize a distributed online implementation for PN anomaly detection, we establish a distributed online augmented Lagrange function and solve it by the alternating direction method of multipliers (ADMM).

Secondly, we propose a CCA-based distributed online PL anomaly detection algorithm, which is realized through analyzing the real-time correlation of measurements between neighbor VNs in respective managers distributedly. To facilitate the real-time analysis, we derive a new method to calculate the canonical correlation vectors, with which the managers only need to store covariance matrices and mean vectors of current VN data instead of all historical data.

Finally, the effectiveness and robustness of the proposed distributed online anomaly detection algorithms are verified on both synthetic and real-world network datasets.

The rest of the paper is organized as follows. Section II presents the related works and Section III describes the system model. The distributed online PN and PL anomaly detection algorithms are elucidated in Section IV and Section V, respectively. Simulation results on both synthetic and real-world network datasets are shown in Section VI. Section VII concludes the paper.

\section{Related Works}
 The key feature of the anomaly detection problem is that, a labeled dataset, where each training sample is labeled as anomalous or normal, is usually prohibitive to obtain \cite{liu2019generative}. Therefore, we usually deal with the anomaly detection as a special classification problem by assuming that the entire dataset contains only one-class samples, i.e., normal samples. OCSVM \cite{scholkopf2001estimating} is a popular unsupervised method to detect anomalies in data. In literature \cite{jiang2013anomaly,schuster2015potentials,9173800,garg2019hybrid}, it has been proved that the OCSVM algorithm has good applicability to a wide range of anomaly detection problems.

CCA algorithm is an important method for analyzing the correlation between two sets of data \cite{jiang2019multimode}. In \cite{chen2017fault,ma2018hierarchical,zhang2019correlation,9080560}, the CCA algorithm has identified whether or not the anomaly exits in the whole system successfully by detecting the variation of correlation between two subsystems, which is also an unsupervised learning method.

However, the classical OCSVM and CCA algorithms are both based on the batch formulation, which requires storing all historical data within a period and trains the normal profiles through offline learning. The process will not only introduce high storage and computation cost but prevent timely detection of anomalies. To conquer this difficulty, online anomaly detection algorithms have attracted much attention in many domains since they can analyze the system data in real-time and enable the timely detection of anomalies. For instance, \cite{xie2018line} proposed a bilateral principle component analysis algorithm to realize the online and accurate traffic anomaly detection for Internet management. In \cite{gomez2011adaptive}, an adaptive OCSVM method was designed to realize online novelty detection for time series scenarios. \cite{he2018admost} proposed an online subspace tracking algorithm to supervise the flight data anomalies of the unmanned aerial vehicle system.  An online and unsupervised anomaly detection algorithm for streaming data was designed in \cite{8917607} using an array of sliding windows and the probability density-based descriptors. These online algorithms showed good anomaly detection performance with relatively short running time and low CPU resources.

Centralized detection, where the training dataset is available in its entirety to one centralized detector, is a well-studied area. However, if the dataset is distributed over more than one location, different approaches need to be taken. Therefore, some distributed algorithms have been designed to achieve the global anomaly detection with datasets distributed in the networks. In \cite{o2016distributed}, a distributed version of principal component analysis algorithm was developed to identify the global anomalies in distributed local datasets. Based on the hierarchical temporal memory, a distributed anomaly detection system was designed in \cite{wang2018distributed} for the security of the in-vehicle network. \cite{9216536} proposed a collaborative intrusion detection framework for Vehicular Ad hoc Networks, which enabled multiple controllers jointly train a global intrusion detection model without direct sub-network flow exchange. In wireless sensor networks, because data used for anomaly detection were distributedly collected, \cite{miao2018distributed} proposed a distributed online one-class support vector machine algorithm for anomaly detection. This algorithm showed a good anomaly detection performance with requiring relatively short running time and low CPU resources. As the virtualized network slices utilize the network resources spanning the whole substrate networks, the training dataset used for PN and PL anomaly detection has a similar distributed property as wireless sensor networks. Therefore, inspired by \cite{miao2018distributed}, we develop a distributed online anomaly detection method that can be applied to the virtualized network slicing environment.


\section{System Model}
\begin{figure*}[htbp]
\centering
\includegraphics[width=5.2in]{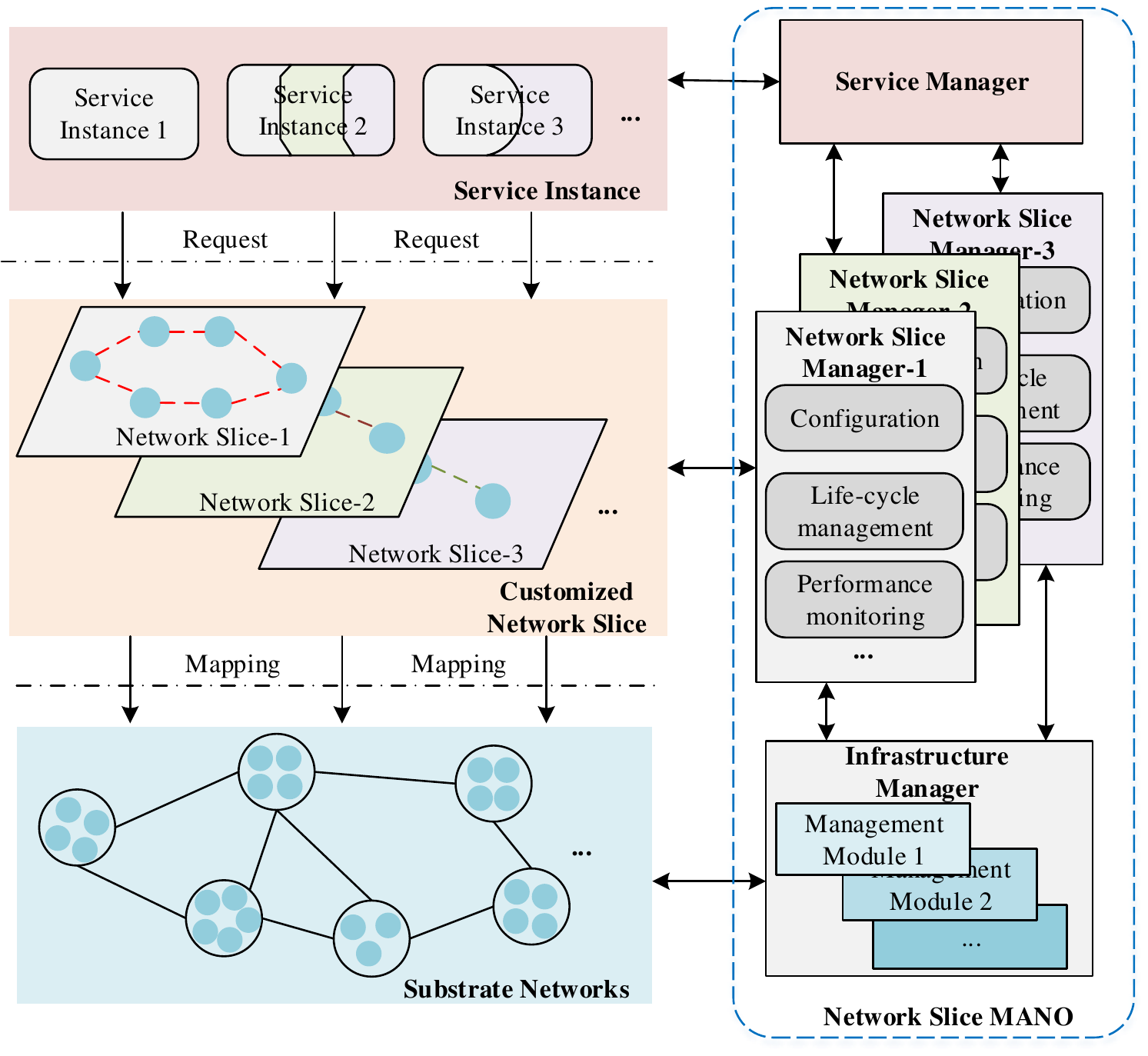}
\caption{Network slice management and orchestration (MANO) overview.}
\end{figure*}

As shown in Fig. 1, when the networks receive a network slice service, its specific demand and required virtual network functions are analyzed by the Service Manager to generate customized SFCs. The Network Slice Manager is responsible for the configuration, life-cycle management and performance monitoring of all VNs in SFCs. During the instantiation process of network slices, multiple SFCs are embedded to a shared substrate network to provide customized services for different slice requests \cite{de20185g}. The Infrastructure Manager takes charge of the management and maintenance of PNs and PLs in substrate networks.

Many unexpected behaviors can cause the anomaly of PNs and PLs in substrate networks, including device malfunction, malicious attackers, fault propagation, external environment changes, etc. Since multiple VNs and VLs from different slices can be mapped to the same PN and PL, one anomalous PN or PL in substrate networks can degrade the performance of multiple network slices. Therefore, the accurate and rapid anomaly detection for PNs and PLs in substrate networks is the prerequisite for ensuring the performance of network slices.

 We present a substrate network by an undirected graph $G=(Q,R)$, where $Q$ and $R$ represent the PN and PL sets contained in the substrate network. Besides, the set of VNs and VLs mapped to PN $q\;(q \in Q)$ and PL $r (r \in R)$ are assumed to be $J_q$ and $J_r$. In the virtualized network slicing environment, the Network Slice Manager will monitor the performance and resource consumption and store the measurements of each customized VN contained in an SFC. VN measurements include processing rate, data flow, queuing delay, processing delay, and the usage of CPU, memory, etc. For each VN $j_q  \in J_q$ mapped to PN $q$, its measurements in a time slot can be represented by ${\bm{x}}_{j_q}=\{x_{j_q}^1,x_{j_q}^2,...,x_{j_q}^p\}$, where $p$ is the number of features. Therefore, the measurements related to the anomaly detection of PN $q$ can be expressed as

\begin{equation}
{\bm{x}}_q  = \left[ {\begin{array}{*{20}c}
   {{\bm{x}}_1 }  \\
   {{\bm{x}}_{\text{2}} }  \\
    \vdots   \\
   {{\bm{x}}_{|J_q |} }  \\
 \end{array} } \right] = \left( {\begin{array}{*{20}c}
   {x_1^1 } & {x_1^2 } &  \ldots  & {x_1^p }  \\
   {x_{\text{2}}^1 } & {x_{\text{2}}^2 } &  \ldots  & {x_{\text{2}}^p }  \\
    \vdots  &  \vdots  &  \ddots  &  \vdots   \\
   {x_{|J_q |}^1 } & {x_{|J_q |}^2 } &  \ldots  & {x_{|J_q |}^p }  \\
 \end{array} } \right),
\end{equation}
where $|J_q|$ represents the number of VNs mapped to PN $q$.

The VN measurements can also be utilized to implement PL anomaly detection. Since the flow passes through each VN of an SFC in sequence, the measurements between neighbor VNs are naturally related. Therefore, we can implement PL anomaly detection by analyzing the correlation of measurements between neighbor VNs, which are mapped to both ends of the PL. The measurements related to the anomaly detection of PL $r$ can be expressed as

\begin{equation}
\begin{gathered}
  {\bm{x}}_m  = \left[ {\begin{array}{*{20}c}
   {{\bm{x}}_1 }  \\
   {{\bm{x}}_{\text{2}} }  \\
    \vdots   \\
   {{\bm{x}}_{|J_r |} }  \\

 \end{array} } \right] = \left( {\begin{array}{*{20}c}
   {x_1^1 } & {x_1^2 } &  \ldots  & {x_1^p }  \\
   {x_{\text{2}}^1 } & {x_{\text{2}}^2 } &  \ldots  & {x_{\text{2}}^p }  \\
    \vdots  &  \vdots  &  \ddots  &  \vdots   \\
   {x_{|J_r |}^1 } & {x_{|J_r |}^2 } &  \ldots  & {x_{|J_r |}^p }  \\

 \end{array} } \right), \hfill \\
  {\bm{x}}_l  = \left[ {\begin{array}{*{20}c}
   {{\bm{x}}_1 }  \\
   {{\bm{x}}_{\text{2}} }  \\
    \vdots   \\
   {{\bm{x}}_{|J_r |} }  \\

 \end{array} } \right] = \left( {\begin{array}{*{20}c}
   {x_1^1 } & {x_1^2 } &  \ldots  & {x_1^d }  \\
   {x_{\text{2}}^1 } & {x_{\text{2}}^2 } &  \ldots  & {x_{\text{2}}^d }  \\
    \vdots  &  \vdots  &  \ddots  &  \vdots   \\
   {x_{|J_r |}^1 } & {x_{|J_r |}^2 } &  \ldots  & {x_{|J_r |}^d }  \\

 \end{array} } \right), \hfill \\
\end{gathered}
\end{equation}
where $m, l\in Q$ represent the both ends of PL $r$, $|J_r|$ is the number of VLs mapped to PL $r$, and $p$ and $d$ denote the number of features for VNs mapped to PN $m$ and $l$, respectively.

 To realize the real-time detection for substrate networks in virtualized network slicing scenarios, the Infrastructure Manager is required to collect VN measurements from the Network Slice Manager sequentially if the centralized detection modes are used. As a central manager, the management module of PN $q$ can utilize measurements of all VNs mapped to it to realize anomalous PN detection. Existing anomaly detection algorithms \cite{Callegari2011A,Kun2017fast,9256316,9064715,jiang2013anomaly,schuster2015potentials,9173800,chen2017fault} usually model all data of a time slot as a long vector and consider it as a single sample in implementing data analysis, so the collected data in a time slot used for the PN anomaly detection can be represented by
\begin{equation}
\begin{gathered}
  {\bm{x}}_q  = \{ {\bm{x}}_1 ,...,{\bm{x}}_{|J_q |} \}  \hfill \\
  \;\;\;\; = \{ \underbrace {x_1^1 ,x_1^2 ,...,x_1^p }_{{\text{VN}}1\;{\text{mapped}}\;{\text{to}}\;{\text{PN}}\;q},\;...,\underbrace {x_{|J_q |}^1 ,x_{|J_q |}^2 ,...,x_{|J_q |}^p }_{{\text{VN }}|J_q |\;{\text{mapped}}\;{\text{to}}\;{\text{PN}}\;q}\} . \hfill \\
\end{gathered}
\end{equation}

Through modeling all measurements of VNs in a time slot mapped to both ends of a PL as two sets of vectors, we can obtain the data used for the PL anomaly detection. The set of VLs mapped to PL $r$ is assumed to be $J_r$, so the anomaly detection for PL $r$ can be realized through analyzing the correlation between the following vectors:

\begin{equation}
\begin{gathered}
  {\bm{x}}_m  = \{ {\bm{x}}_1 ,...,{\bm{x}}_{|J_r |} \}  \hfill \\
  \;\;\;\; = \{ \underbrace {x_1^1 ,x_1^2 ,...,x_1^p }_{{\text{VN}}1\;{\text{mapped}}\;{\text{to}}\;{\text{PN}}\;m},...,\underbrace {x_{_{|J_r |} }^1 ,x_{_{|J_r |} }^2 ,...,x_{_{|J_r |} }^p }_{{\text{VN }}_{|J_r |} \;{\text{mapped}}\;{\text{to}}\;{\text{PN}}\;m}\}  \hfill \\
  {\bm{x}}_l  = \{ {\bm{x}}_1 ,...,{\bm{x}}_{|J_r |} \}  \hfill \\
  \;\;\;\; = \{ \underbrace {x_1^1 ,x_1^2 ,...,x_1^d }_{{\text{VN}}1\;{\text{mapped}}\;{\text{to}}\;{\text{PN}}\;l},...,\underbrace {x_{_{|J_r |} }^1 ,x_{_{|J_r |} }^2 ,...,x_{_{|J_r |} }^d }_{{\text{VN }}_{|J_r |} \;{\text{mapped}}\;{\text{to}}\;{\text{PN}}\;l}\}.  \hfill \\
\end{gathered}
\end{equation}

However, the centralized detection modes will introduce additional communication and storage cost by collecting all VN measurements to the Infrastructure Manager. Besides, the data privacy of different slices can be compromised. Therefore, to determine whether or not the anomaly exists in PNs and PLs rapidly and accurately with low communication and storage cost, the distributed online PN and PL anomaly detection methods are required to analyze the real-time measurements of VNs in respective Network Slice Manager distributedly.

\textit{Notation}:  In this paper, boldface letters are used to denote vectors and $( \cdot )^{\rm{T}}$ is the operator of transposition. $|| \cdot ||$ represents the Euclidean norm. $\bm{0}$ and $\bm{1}$ denote the vectors of all zeros and ones, respectively. Other notations will be defined if necessary.

\section{Distributed Online PN Anomaly Detection Based on Decentralized OCSVM}
From the perspective of the centralized mode, to implement the anomaly detection for a PN, its management module will model measurements of all VNs mapped to it within a period as a series of vectors to form the training set. For each VN $j_q  \in J_q$, its training set can be expressed as $S_{j_q }  = \{ ({\bm{x}}_{j_q n} ,y_{j_q n} ),n = 1,2,....,N_{j_q } \}$, where ${\bm{x}}_{j_q n}  \in \mathbb{R}^p$, $N_{j_q }$ is the number of training data, and $y_{j_q n}  =  + 1$ indicates that all training data are normal samples. As the training set of VN is $S_{j_q }$, the collected data of PN $q$ within a period can be represented by $S_q {\rm{ = }}\bigcup\nolimits_{j_q  \in J_q } {S_{j_q } }= \{ ({\bm{x}}_{qn} ,y_{qn} ),n = 1,2,...,N_q \}$, where ${\bm{x}}_{qn}  = \bigcup\nolimits_{j_q  \in J_q } {{\bm{x}}_{j_q n} }$ and $y_{qn}  = y_{1n}  = ,..., = y_{j_q n}  =  + 1$. Generally, the training data are complex and linearly inseparable. Therefore, OCSVM algorithm uses the feature mapping function to map the training data from input space to high-dimensional feature space. The purpose of the mapping is to make the data linearly separable in the new space. Through finding a hyperplane in the feature space, OCSVM algorithm can isolate the mapped samples from the origin with maximum margin \cite{scholkopf2001estimating}.

We use the feature mapping function $\bm{\phi} ( \cdot )$ to map the input data $\bm{x}_q$ into a reproducing kernel Hilbert space, and then calculate the inner product of the feature mapping function by the kernel function $k(\bm{x}_{q1} ,\bm{x}_{q2} )$. The commonly used kernel function is the Gaussian kernel function \cite{scholkopf2001estimating}
\begin{equation}
\begin{gathered}
k({\bm{x}}_{q1} ,{\bm{x}}_{q2} ) = \langle {\bm{\phi }}({\bm{x}}_{q1} ),{\bm{\phi }}({\bm{x}}_{q2} )\rangle
   \hfill \\
\;\;\;\;\;\;\;\;\;\;\;\;\;\;\;\;\;\;  = \exp \left[ { - ||{\bm{x}}_{q1}  - {\bm{x}}_{q2} ||^2 /\sigma ^2} \right],
 \end{gathered}
\end{equation}
where $\sigma$ denotes the kernel width. According to formula (1), we can get that $k({\bm{x}}_q ,{\bm{x}}_q ) =  < {\bm{\phi }}({\bm{x}}_q ),{\bm{\phi }}({\bm{x}}_q ) >  = 1$, which means that the mapping function $\bm{\phi} ( \cdot )$ will map the original samples into a hypersphere with  center $\bm{0}$ and radius 1. As
\begin{equation}
\begin{gathered}
 k({\bm{x}}_{q1} ,{\bm{x}}_{q2} ) = \exp \left[ { - ||{\bm{x}}_{q1}  - {\bm{x}}_{q2} ||^2 /\sigma ^2 } \right]
   \hfill \\
\;\;\;\;\;\;\;\;\;\;\;\;\;\;\;\;\;\; = ||{\bm{\phi }}({\bm{x}}_{q1} )|| \cdot ||{\bm{\phi }}({\bm{x}}_{q2} )|| \cdot \cos \langle {\bm{\phi }}({\bm{x}}_{q1} ),{\bm{\phi }}({\bm{x}}_{q2} )\rangle,
 \end{gathered}
\end{equation}
if sample ${\bm{x}}_{q1}$ is normal while ${\bm{x}}_{q2}$ is distant from ${\bm{x}}_{q1}$ in input space, the value of $k({\bm{x}}_{q1} ,{\bm{x}}_{q2} )$ will be small, which means that the value of $ \cos \langle {\bm{\phi }}({\bm{x}}_{q1} ),{\bm{\phi }}({\bm{x}}_{q2} )\rangle$ is small and the angle between ${\bm{\phi }}({\bm{x}}_{q1} )$ and ${\bm{\phi }}({\bm{x}}_{q2} )$ is large. In this case, we usually consider ${\bm{x}}_{q2}$ as anomalous data, which can be isolated from normal data through finding ($\bm{w}_q$, $\rho _q$), which are a weight vector and an offset parameterizing a hyperplane ${\bm{w}}_q^{\rm{T}} {\bm{\phi }}({\bm{x}}) - \rho _q  = 0$ in the feature space associated with the kernel. The hyperplane can isolate the training samples from the origin with maximum margin $(\rho _q )/(||{\bm{w}}_q ||)$.

Based on the above reasoning, the problem of the anomaly detection for PN $q$ can be formulated as the following optimization objective:
\begin{equation}
\begin{gathered}
\mathop {\min }\limits_{{\bm{w}}_q ,\rho _q } \;\frac{1}{2}||{\bm{w}}_q^{\rm{T}} ||^2  + C\sum\limits_{n = 1}^{N_q} {\xi _{qn} }  - \rho _q
   \hfill \\
{\rm{s.t.}}\; {\bm{\phi }}^{\rm{T}}({\bm{x}}_{qn} ){\bm{w}_q} \ge \rho _q  - \xi _{qn} ,\;\;\;\;n = 1,2,....,N_q,
    \hfill \\
\;\;\;\;\;\;\xi _{qn}  \ge 0,\;\;\;\;\;\;\;\;\;\;\;\;\;\;\;\;\;\;\;\;\;\;\;\;\;\;n = 1,2,....,N_q, \;\;\;\;\;\;
\end{gathered}
\end{equation}
where $C$ is the penalty parameter, and $\xi _{qn}$ denotes the slack variable that allows some samples to appear in the margin of the hyperplane. After obtaining the optimal ${\bm{w}}_q^*$ and $\rho_q ^{\rm{*}}$ for anomaly detection, we can define the discriminant function as
\begin{equation}
g({\bm{x}}_{q n} ) = {\mathop{\rm sgn}} ({\bm{\phi}}^{\rm{T}}({\bm{x}}_{q n} ){{\bm{w}_q}^{\rm{*}}} - \rho_q ^* ),\;\forall q  \in Q ,\;n \in N_{q }.
\end{equation}

For any data ${\bm{x}}_{q n}$, it can be determined as a normal data if $g({\bm{x}}_{q n} ) \ge 0$, and an anomalous data otherwise.
\subsection{Decentralization of OCSVM}
The above-mentioned centralized detection mode requires the Infrastructure Manager collecting measurements of all VNs sequentially from the Network Slice Manager and analyze them as a whole, which will not only bring high communication and storage cost but compromise the data privacy of different slices. Therefore, we distribute the OCSVM-based PN anomaly detection problem to the Network Slice Manager instead of transmitting all VN measurements to the Infrastructure Manager. The Network Slice Manager is responsible to analyze the measurements of each VN.

To present the centralized problem (7) in a distributed form, a group of decentralized quadratic programming problems that satisfy the consensus constraints are established \cite{forero2010consensus}.

 The global variables $\{ {\bm{w}}_q ,\rho _q \}$ are replaced by auxiliary local variables $\{ {\bm{w}}_{j_q } ,\rho _{j_q } \} _{j_q  = 1}^{J_q }$ of each VN. To ensure the consistency among local variables, extra consensus constraints are added to the new objective. Therefore, the distributed form of (7) can be expressed as
\begin{equation}
\begin{gathered}
 \mathop {\min }\limits_{{\{\bm{w}}_{j_q} ,\rho _{j_q }\}} \;\frac{1}{2}\sum\limits_{j_q  = 1}^{|J_q |} {||{\bm{w}}_{j_q } ||^2 }  + |J_q |C\sum\limits_{j_q  = 1}^{|J_q |} {\sum\limits_{n = 1}^{N_{j_q } } {\xi _{j_q n} } }  - \sum\limits_{j_q  = 1}^{|J_q |} {\rho _{j_q } }
  \hfill \\
 \;{\rm{s.t.}}\; \bm{\phi} ^{\rm{T}}({\bm{x}}_{j_q n} ){\bm{w}}_{j_q } \ge \rho _{j_q }  - \xi _{j_q n} ,\;\;\forall j_q  \in J_q ,\;n = 1,2,....,N_{j_q },
  \hfill \\
 \;\;\;\;\;\;\;\xi _{j_q n}  \ge 0,\;\;\;\;\;\;\;\;\;\;\;\;\;\;\;\;\;\;\;\;\;\;\;\;\;\;\forall j_q  \in J_q ,\;n = 1,2,....,N_{j_q },
  \hfill \\
 \;\;\;\;\;\;\;{\bm{w}}_{j_q }  = \;{\bm{w}}_{i_q } ,\rho _{j_q }  = \rho _{i_q } ,\;\;\;\;\;\;\;\;\forall j_q ,i_q  \in J_q. \;\;\;\;\;\;\;\;\;\;\;\;\;\;\;\;\;\;\;\;\;\;\;\;
\end{gathered}
\end{equation}

As the local variables $\{ {\bm{w}}_{j_q } ,\rho _{j_q } \} _{j_q  = 1}^{J_q }$ of VNs mapped to the same PN satisfy ${\bm{w}}_1  = {\bm{w}}_2  = ,..., = {\bm{w}}_{J_q }  = {\bm{w}}_q$ and $\rho _1  = \rho _2  = ,..., = \rho _{J_q }  = \rho _q$, we can rewrite (9) as
\begin{equation}
\begin{gathered}
\mathop {\min }\limits_{{\{\bm{w}}_q ,\rho _q\}} \;|J_q |\;\left( {\frac{1}{2}||{\bm{w}}_q ||^2  + C\sum\limits_{j_q  = 1}^{|J_q |} {{\bm{1}}_{j_q }^{\rm{T}} {\bm{\xi }}_{j_q } }  - \rho _q } \right)
 \hfill \\
 \;{\rm{s.t.}}\; {\bm{\Phi}} ^{\rm{T}}({\bm{X}}_{j_q } ){\bm{w}}_q \ge \rho _q {\bm{1}}_{j_q }  - {\bm{\xi }}_{j_q } ,\;\;\;\;\forall j_q  \in J_q,
 \hfill \\
 \;\;\;\;\;\;\;\;{\bm{\xi }}_{j_q }  \ge \bm{0}_{j_q } ,\;\;\;\;\;\;\;\;\;\;\;\;\;\;\;\;\;\;\;\;\;\;\;\;\;\;\;\;\;\forall j_q  \in J_q, \;\;
\end{gathered}
\end{equation}
where $\bm{\Phi} {\rm{(}}{\bm{X}}_{j_q } {\rm{) = [}}\bm{\phi} ({\bm{x}}_{j_q 1} ),...,\bm{\phi}({\bm{x}}_{j_q N_{j_q } } ){\rm{]}}$ and $\{ {\bm{w}}_q ,\rho _q \}$  is the feasible solution for (7). Since the factor $|J_q |$ is a constant, the objective (10) is equivalent to (7). Therefore, the distributed objective (9) is also equivalent to (7).

Similar to the objective (7), $k({\bm{x}}_{j_q n} ,{\bm{x}}_{i_q m} ) = \langle \bm{\phi} ({\bm{x}}_{j_q n} ),\bm{\phi} ({\bm{x}}_{i_q m} )\rangle ,\;\forall j_q ,i_q  \in J_q,\;n\in N_{j_q },\;m\in N_{i_q },$ should be calculated instead of $\bm{\phi} ({\bm{x}}_{j_q n} )$ or $\bm{\phi} ({\bm{x}}_{i_q m} )$ in solving the distributed problem (9), where $\bm{\phi} ({\bm{x}}_{j_q n} )$ and $\bm{\phi} ({\bm{x}}_{i_q m} )$ are unknown \cite{miao2018distributed}. Since the measurements are distributed in many VNs, it is difficult to calculate $k({\bm{x}}_{j_q n} ,{\bm{x}}_{i_q m} )$ without information exchange among VNs. Therefore, the random approximation method \cite{rahimi2008random} is adopted to approximate $\bm{\phi} ({\bm{x}}_{j_q n} )$ with $\bm{z} ({\bm{x}}_{j_q n} )$.

Through introducing a random approximation function ${\bm{z}}:\mathbb{R}^p  \to \mathbb{R}^D$, we can map the input data to a random feature space, where $D$ is the dimension of the random feature space and satisfies $D > p$. Using this technique, $\bm{\phi} ({\bm{x}}_{j_q n} )$ can be approximated by $\bm{z} ({\bm{x}}_{j_q n} )$. The inner product calculation between $\bm{\phi} ({\bm{x}}_{j_q n} )$ and $\bm{\phi} ({\bm{x}}_{i_q m} )$ can be expressed as
\begin{equation}
k({\bm{x}}_{j_q n} ,{\bm{x}}_{i_q m} ) = \langle \bm{\phi} ({\bm{x}}_{j_q n} ),{\bm{\phi }}({\bm{x}}_{i_q m} )\rangle  \approx {\bm{z}}({\bm{x}}_{j_q n} )^{\rm{T}} {\bm{z}}({\bm{x}}_{i_q m} ),
\end{equation}
where ${\bm{z}}({\bm{x}}_{j_q n} ) = {\rm{[}}z_{{\bm{\omega }}_1 } {\rm{(}}{\bm{x}}_{j_q n} {\rm{),}}...{\rm{,}}z_{{\bm{\omega }}_D } {\rm{(}}{\bm{x}}_{j_q n} {\rm{)]}}^{\rm{T}}$, and ${\bm{z}}_{{\bm{\omega }}_i } ({\bm{x}}_{j_q n} )$ is a mapping function and given by
\begin{equation}
\begin{gathered}
\bm{z}_{{\bm{\omega }}_{\bm{i}} } {\rm{(}}{\bm{x}}_{j_q n} {\rm{)}} = \sqrt {\frac{2}{D}} \cos ({\bm{\omega }}_i^{\rm{T}} {\bm{x}}_{j_q n}  + \vartheta _i ),\;\;
 \hfill \\
 \;\;\;\;\;\;\;\;\;\;\;\;\;\;\;\;\;\;\;\;\;\;\;\;i = 1,2,...,D,\;\forall j_q  \in J_q ,\;n \in N_{j_q },
  \hfill \\
 \end{gathered}
\end{equation}
where $\vartheta _i$ is uniformly drawn from $[0,2\pi ]$ and ${\bm{\omega }}_i$ is drawn from $p({\bm{\omega }}) = (2\pi )^{ - (D/2)} e^{ - [||{\bm{\omega }}||^2 /2]}$, which denotes the Fourier transform of Gaussian kernel function \cite{rahimi2008random}.

Note that the random approximate function $\bm{z}({\bm{x}}_{j_q n} )$ and the mapping feature function ${\bm{\phi }}({\bm{x}}_{j_q n} )$ share the same properties in reproducing kernel Hilbert space.  $\bm{z}({\bm{x}}_{j_q n} )$ also can map the original data into a hypersphere with center $\bm{0}$ and radius 1 similar to ${\bm{\phi }}({\bm{x}}_{j_q n} )$. Besides, as proved in \cite{rahimi2008random}, if the dimension of $\bm{z}({\bm{x}}_{j_q n} )$ is proper, $k({\bm{x}}_{j_q n} ,{\bm{x}}_{i_q m} )$ can be well approximated by ${\bm{z}}({\bm{x}}_{j_q n} )^{\rm{T}} {\bm{z}}({\bm{x}}_{i_q m} )$. Then, we have
\begin{equation}
\begin{gathered}
 k({\bm{x}}_{j_q n} ,{\bm{x}}_{i_q m} )
  \hfill \\
 = ||{\bm{\phi }}({\bm{x}}_{j_q n} )|| \cdot ||{\bm{\phi }}({\bm{x}}_{i_q m} )|| \cdot \cos \langle {\bm{\phi }}({\bm{x}}_{j_q n} ),{\bm{\phi }}({\bm{x}}_{i_q m} )\rangle
   \hfill \\
 \approx ||{\bm{z}}({\bm{x}}_{j_q n} )|| \cdot ||{\bm{z}}({\bm{x}}_{i_q m} )|| \cdot \cos \langle {\bm{z}}({\bm{x}}_{j_q n} ),{\bm{z}}({\bm{x}}_{i_q m} )\rangle,
 \end{gathered}
\end{equation}
which means that $\cos \langle {\bm{\phi }}({\bm{x}}_{j_q n} ),{\bm{\phi }}({\bm{x}}_{i_q m} )\rangle  \approx \cos \langle {\bm{z}}({\bm{x}}_{j_q n} ),{\bm{z}}({\bm{x}}_{i_q m} )\rangle$. Therefore, if data ${\bm{x}}_{j_q n}$ is close to ${\bm{x}}_{i_q m}$, their corresponding mapping vectors ${\bm{z}}_{j_q n}$ and ${\bm{z}}_{i_q m}$ will be close in the new random feature space.

Based on the above reasoning, the problem (9) can be transformed as follows:
\begin{equation}
\begin{gathered}
\mathop {\min }\limits_{{\{\bm{w}}_{j_q} ,\rho _{j_q }\}} \; \sum\limits_{j_q  = 1}^{J_q } {\left( {\;\frac{1}{2}||{\bm{w}}_{j_q } ||^2 {\rm{ + |}}J_q |C\sum\limits_{n = 1}^{N_{j_q } } {\xi _{j_q n} }  - \rho _{j_q } } \right)}
   \hfill \\
 \;{\rm{s.t.}}\; {\bm{z}}^{\rm{T}}({\bm{x}}_{j_q n} ){\bm{w}}_{j_q } \ge \rho _{j_q }  - \xi _{j_q n} ,\;\forall j_q  \in J_q ,\;n = 1,2,....,N_{j_q },
   \hfill \\
 \;\;\;\;\;\;\xi _{j_q n}  \ge 0,\;\;\;\;\;\;\;\;\;\;\;\;\;\;\;\;\;\;\;\;\;\;\;\;\;\;\forall j_q  \in J_q ,\;n = 1,2,....,N_{j_q },
   \hfill \\
 \;\;\;\;\;\;{\bm{w}}_{j_q }  = \;{\bm{w}}_{i_q } ,\rho _{j_q }  = \rho _{i_q } ,\;\;\;\;\;\;\; \forall j_q ,i_q  \in J_q,
  \hfill \\
 \end{gathered}
\end{equation}
where $ {\bm{x}}_{j_q n}  \in S_{j_q }$ denotes the training samples of VN $j_q$. It should be noted that the symbol ${\bm{w}}_{j_q }$ in (14) has different dimension with that in (9), but for simplicity, we choose to use the same symbol to represent them.

\subsection{Distributed Online PN Anomaly Detection}
For each VN, new unlabeled measurements will be generated at each time. If we keep all historical data of each VN within a period for offline training, it will not only introduce high storage and computation cost but prevent timely detection of anomalies. Therefore, unlabeled training data are expected to be processed in an online mode. Based on the above analysis, we propose a distributed online PN anomaly detection algorithm based on the decentralized OCSVM.

Through introducing the Lagrange multipliers $\{ \kappa _j \}$, $\{ \lambda _j \}$, $\{ \bm{\alpha} _{ji} \}$ and $\{ \beta _{ji} \}$, the augmented Lagrange function for the problem (14) is expressed as

\begin{equation}
\begin{gathered}
L(\{ {\bm{w}}_{j_q } \} ,\{ \rho _{j_q } \} ,\{ \xi _{j_q n} \} ,\{ \kappa _j \} ,\{ \lambda _j \} ,\{ \bm{\alpha} _{ji} \} ,\{ \beta _{ji} \} ) =
 \hfill \\
  \sum\limits_{j_q  = 1}^{|J_q |} {\left( \begin{gathered}
  \;\frac{1}
{2}||{\bm{w}}_{j_q } ||^2  + |J_q |C\sum\limits_{n = 1}^{N_{j_q } } {\xi _{j_q n} }  - \rho _{j_q }  - \frac{{\kappa _j }}
{{N_{j_q } }}\sum\limits_{n = 1}^{N_{j_q } } {\xi _{j_q n} }  \hfill \\
   - \frac{{\lambda _j }}
{{N_{j_q } }}\sum\limits_{n = 1}^{N_{j_q } } {\left[ {{\bm{z}}^{\rm{T}}({\bm{x}}_{j_q n} ){\bm{w}}_{j_q }  - \rho _{j_q }  + \xi _{j_q n} } \right] }  \hfill \\
 + \sum\limits_{i_q  = 1 }^{|J_q |} {\bm{\alpha} _{ji}^{\rm{T}}({\bm{w}}_{j_q }  - {\bm{w}}_{i_q })  + } \sum\limits_{i_q  = 1}^{|J_q |}{\beta _{ji} (\rho _{j_q }  - \rho _{i_q } )}  \hfill \\
   + \frac{\eta }
{2}\sum\limits_{i_q  = 1 }^{|J_q |} {[||{\bm{w}}_{j_q }  - {\bm{w}}_{i_q } ||^2  + ||\rho _{j_q }  - \rho _{i_q } ||^2 ]}  \hfill \\
\end{gathered}  \right)}.  \hfill \\
\end{gathered}
\end{equation}

 Here, the last term is the regularization term, which plays two roles: (1) It eliminates the condition that $L( \bm{\theta} _j )$ must be differentiable, where $\bm{\theta} _j$ represents the set of variables;  (2) The convergence speed of ADMM could be controlled by adjusting the augmented Lagrange parameter $\eta$. Then, $L( \bm{\theta} _j )$ could be minimized in a cycle fashion: at each iteration, we minimize $L( \bm{\theta} _j )$ with respect to one variable while keeping all other variables fixed \cite{forero2010consensus}. The ADMM solution at each iteration $t+1$ takes the form
\begin{subequations}
\begin{equation}
\begin{gathered}
  \{ {\bm{w}}_{j_q } (t + 1),\rho _{j_q } (t + 1),\xi _{j_q n} (t + 1)\}  =  \hfill \\
\mathop {\arg \min }\limits_{\{ {\bm{w}}_{j_q } \} ,\{ \rho _{j_q } \} ,\{ \xi _{j_q n} \} } L\left( \begin{gathered}\{ {\bm{w}}_{j_q } \} ,\{ \rho _{j_q } \} ,\{ \xi _{j_q n} \} ,\{ \kappa _j \} \hfill \\
\{ \lambda _j \} , \bm{\alpha} _{ji}(t) \} ,\{ \beta _{ji}(t) \} \end{gathered}  \right), \hfill \\
\end{gathered}
\end{equation}
\begin{equation}
\bm{\alpha} _{j} (t + 1) = \bm{\alpha} _{j}(t) + \frac{\eta }
{2}\sum\limits_{i_q  = 1}^{|J_q |} {({\bm{w}}_{j_q } (t + 1) - {\bm{w}}_{i_q } (t + 1))},
\end{equation}
\begin{equation}
\beta _{j} (t + 1) = \beta _{j} (t) + \frac{\eta }
{2}\sum\limits_{i_q  = 1}^{|J_q |} {(\rho _{j_q } (t + 1) - \rho _{i_q } (t + 1))},
\end{equation}
\end{subequations}
where $\bm{\alpha} _j (t) = \sum\nolimits_{i = 1}^{|J_q |} {\bm{\alpha} _{ji} (t)}$ and $\beta _j (t) = \sum\nolimits_{i = 1}^{|J_q |} {\beta _{ji} (t)}$.

Model parameters $\{ {\bm{w}}_{j_q } \}$, $\{ \rho _{j_q } \}$ and $\{ \xi _{j_q n} \}$ of decentralized OCSVM could be obtained by solving the problem (16a). It is obvious that the problem (16a) is a batch formulation of distributed algorithm, where all data of each VN are required to be available. This will bring a big challenge to the storage resources of the networks. To overcome this problem, a distributed online augmented Lagrange function is defined in equation (17) by replacing $\rho _{i_q }$ and ${\bm{w}}_{i_q }$  at time $t$ with $\frac{1}{2}(\rho _{j_q } (t) + \rho _{i_q } (t))$ and $\frac{1}{2}({\bm{w}}_{j_q } (t) + {\bm{w}}_{i_q } (t))$ as in \cite{forero2010consensus}.
\newcounter{TempEqCnt} 
\setcounter{TempEqCnt}{\value{equation}} 
\setcounter{equation}{16} 
\begin{figure*}[ht] 
	\begin{equation}
\begin{gathered}
  L'\left( {\{ {\bm{w}}_{j_q } \} ,\{ \rho _{j_q } \} ,\xi _{j_q } (t),\{ \kappa _j \} ,\{ \lambda _j \}, \{ {\bm{w}}_{j_q } (t)\} ,\{ \rho _{j_q } (t)\} ,\{ \bm{\alpha} _j (t)\} ,\{ \beta _j (t)\} } \right) =  \hfill \\
  \sum\limits_{j_q  = 1}^{|J_q |} {\left( \begin{gathered}
  \;\frac{1}
{2}||{\bm{w}}_{j_q } ||^2  + |J_q |C\xi _{j_q } (t) - \rho _{j_q }  - \kappa _j \xi _{j_q } (t) - \lambda _j [{\bm{z}}^{\rm{T}}_{j_q } (t){\bm{w}}_{j_q }  - \rho _{j_q }  + \xi _{j_q } (t)] + 2\bm{\alpha} _j^{\rm{T}} (t){\bm{w}}_{j_q }  \hfill \\
   + 2\beta _j (t)\rho _{j_q }  + \frac{\eta }
{2}\sum\limits_{i_q  = 1}^{|J_q |} {[||{\bm{w}}_{j_q }  - \frac{1}
{2}({\bm{w}}_{j_q } (t) + {\bm{w}}_{i_q } (t))||^2  + ||\rho _{j_q }  - \frac{1}
{2}(\rho _{j_q } (t) + \rho _{i_q } (t))||^2 ]}  \hfill \\
\end{gathered}  \right)}. \hfill \\
\end{gathered}
	\end{equation}
\end{figure*}

 In (17), $t$ is the time instant of online learning, ${\bm{z}}_{j_q } (t) = {\rm{[}}z_{{\bm{\omega }}_1 } ({\bm{x}}_{j_q } (t)){\rm{,}}...{\rm{,}}z_{{\bm{\omega }}_D } ({\bm{x}}_{j_q } (t)){\rm{]}}^{\rm{T}}$, where ${\bm{x}}_{j_q } (t)$ is the training data of VN $j_q$ at time $t$. $\xi _{j_q } (t)$ is the slack variable for ${\bm{x}}_{j_q } (t)$.

From the KKT conditions for (17) it follows that
\begin{subequations}
\begin{equation}
{\bm{w}}_{j_q } (t + 1) = \frac{1}{A} \left( \begin{gathered}
  \bm{z}_{j_q } (t + 1)\lambda _j (t + 1) - 2\bm{\alpha} _j (t) \hfill \\
   + \frac{\eta }
{2}\sum\limits_{i_q  = 1}^{|J_q |} {({\bm{w}}_{j_q } (t) + {\bm{w}}_{i_q } (t))} \hfill \\
\end{gathered}  \right),
\end{equation}
\begin{equation}
\rho _{j_q } (t + 1) = \frac{1}{A-1} \left( \begin{gathered}
  1 - \lambda _j (t + 1) - 2\beta _j (t) \hfill \\
   + \frac{\eta }
{2}\sum\limits_{i_q  = 1}^{|J_q |} {(\rho _{j_q } (t) + \rho _{i_q } (t))}  \hfill \\
\end{gathered}  \right),
\end{equation}
\begin{equation}
0  = |J_q |C  - \lambda _j  - \kappa _j,
\end{equation}
\end{subequations}
where $A = \eta |J_q |+1$. The KKT conditions require $\lambda _j  \geqslant 0 ,\;\kappa _j  \geqslant 0$, so (18c) is allowed to be replaced by $0  \leqslant \lambda _j  \leqslant |J_q |C$. To implement the update (18a) and (18b) for every VN, the optimal Lagrange multipliers $\{ \lambda _j (t + 1)\} $ are obtained by solving the dual problem of (17).  The corresponding dual function is given by
\begin{equation}
\begin{gathered}
  L'_\lambda  \{ \lambda _j (t + 1)\} =  \hfill \\
   - [\bm{z}_{j_q }^{\rm{T}} (t + 1)A ^{ - 1} \bm{z}_{j_q } (t + 1) + (A-1)^{ - 1} ]\lambda^2 _j  \hfill \\
   + [A ^{ - 1} l_j (t) + (A-1)^{ - 1} (1 - h_j (t))]\lambda _j,  \hfill \\
\end{gathered}
\end{equation}
where $l_j (t) = 2\bm{\alpha} _j (t) - \frac{\eta }{2}\sum\limits_{i_q  = 1}^{|J_q |} {({\bm{w}}_{j_q } (t) + {\bm{w}}_{i_q } (t))}$ and $h_j (t) = 2\beta _j (t) - \frac{\eta }{2}\sum\limits_{i_q  = 1}^{|J_q |} {(\rho _{j_q } (t) + \rho _{i_q } (t))}$. $\{ \lambda _j (t + 1)\} $ is given by
\begin{equation}
\begin{gathered}
  \lambda _j {\text{(}}t + 1{\text{)}} =
\mathop {\arg \max }\limits_{0  \leqslant \lambda _j  \leqslant |J_q |C }   -  [\bm{z}_{j_q }^{\rm{T}} (t + 1)A ^{ - 1} \bm{z}_{j_q } (t + 1)+ \hfill \\
 (A-1)^{ - 1} ]\lambda ^2_j
   + [A ^{ - 1} l_j (t) + (A-1)^{ - 1} (1 - h_j (t))]\lambda _j.  \hfill \\
\end{gathered}
\end{equation}

Algorithm 1 presents the detailed steps of the distributed online PN anomaly detection algorithm. In Algorithm 1, steps (2)-(8) are implemented in Network Slice Manager and steps (9)-(13) are implemented in Infrastructure Manager. Since the distributed online algorithm can eliminate the impact of anomalous data on estimates in each iteration (step 12), it can hold high detection accuracy without any labeled data.

\begin{algorithm}[t]
\caption{OCSVM-based distributed online PN anomaly detection algorithm}
\begin{algorithmic}[1]
\REQUIRE Initialize the iteration number $T$, dimension $D$, multipliers $\bm{\alpha}_j(0)$ and $\beta_j(0)$, and estimates ${\bm{w}}_{j_q } (0)$ and $\rho _{j_q } (0)$  of each VN
\FOR {$t = 1,2,...,T$}
\FOR {$j_q  \in J_q$}     
\STATE VN $j_q$ obtains a new sample ${\bm{x}}_{j_q } (t)$, and calculate the approximation ${\bm{z}}_{j_q } (t)$ of $\bm{\phi} (\bm{x}_{j_q } (t))$ by the random approximate function
\STATE Compute $ \lambda _j (t) $, ${\bm{w}}_{j_q } (t)$ and $\rho _{j_q } (t)$ according to (20), (18a) and (18b)
\STATE Compute $ \bm{\alpha} _j (t) $ and $\beta _{j_q } (t)$ according to (16b) and (16c)
\STATE Compute $g({\bm{x}}_{j_q } (t)) = {\mathop{\rm sgn}} ( {\bm{z}}_{j_q }^{\rm{T}} (t){\bm{w}}_{j_q } (t) - \rho _{j_q } (t))$
\STATE Send $g({\bm{x}}_{j_q } (t))$, ${\bm{w}}_{j_q } (t)$ and $\rho _{j_q } (t)$ to the management module of PN $q$
\ENDFOR
\IF {$\prod\nolimits_{j_q  \in J_q } {g({\bm{x}}_{j_q } (t))}  =  = 1$} 
\STATE  PN $q$ is detected as normal at time $t$, and update estimates ${\bm{w}}_{j_q } (t)$ and $\rho _{j_q } (t)$.  Then, broadcast them to related VNs
\ELSE
\STATE  PN $q$ is detected as anomaly at time $t$, and reserve estimates at time $t-1$ and discard current ones
\ENDIF
\ENDFOR
\end{algorithmic}
\end{algorithm}
\addtolength{\topmargin}{0.01in}

 From formulas (18a) and (18b), we can know that VN $j_q$ requires estimates ${\bm{w}}_{i_q } (t)$ and $\rho _{i_q } (t)(i_q  \in J_q)$ to be available when updating estimates ${\bm{w}}_{j_q } (t+1)$ and $\rho _{j_q } (t+1)$. According to \cite{scholkopf2001estimating}, the time complexity of classical OCSVM algorithm is $O(N_q^3 )$. Since all training samples and their labels need to be available in OCSVM, its storage complexity is $O(N_q|J_q |(p+1))$. The calculation of the online OCSVM algorithm depends on the number of iterations $T$ and the number of VNs $|J_q|$ mapped to PN $q$, so its time complexity is $O(N_q|J_q|)$ when the number of iterations $T = N_q$. Since the online algorithm only needs current measurements to be available, its storage complexity is $O(|J_q |p)$. Due to the distributed online mode, the time and storage complexity of the proposed PN anomaly detection algorithm are $O(N_q)$ and $O(p)$, respectively.

\section{CCA-based distributed online PL Anomaly Detection}
The basic principle of PL anomaly detection: Since the flow passes through each VN of an SFC in sequence, the measurements between neighbor VNs are naturally related \cite{cotroneo2017fault}. Assume that the virtual path ${\text{VN}}_m \mathop  \to \limits^{{\text{VL}}_{m,l} } {\text{VN}}_{l}$ is instantiated into the path ${\text{PN}}_m \mathop  \to \limits^{{\text{PL}}_{m,l} } {\text{PN}}_{l}$, where virtual link ${\text{VL}}_{m,l}$ is instantiated into physical link ${\text{PL}}_{m,l}$, and ${\rm{VN}}_{m}$ and ${\rm{VN}}_{l}$ are instantiated into ${\rm{PN}}_m$ and ${\rm{PN}}_{l}$, respectively. Then, the correlation of measurements between ${\rm{VN}}_m$ and ${\rm{VN}}_{l}$ will be stable within a certain range if ${\rm{PL}}_{m,l}$ is in a normal state, and will change if an anomaly occurs in ${\rm{PL}}_{m,l}$. As shown in Fig. 2, if the working state of PL 34 becomes anomalous, the correlation of measurements between VN 2 and 3 in Slice 2 will change. Therefore, we can implement PL anomaly detection based on the correlation of measurements between neighbor VNs, which are mapped to both ends of this PL.

\begin{figure}[t]
 \centering
\includegraphics[width=3.5in]{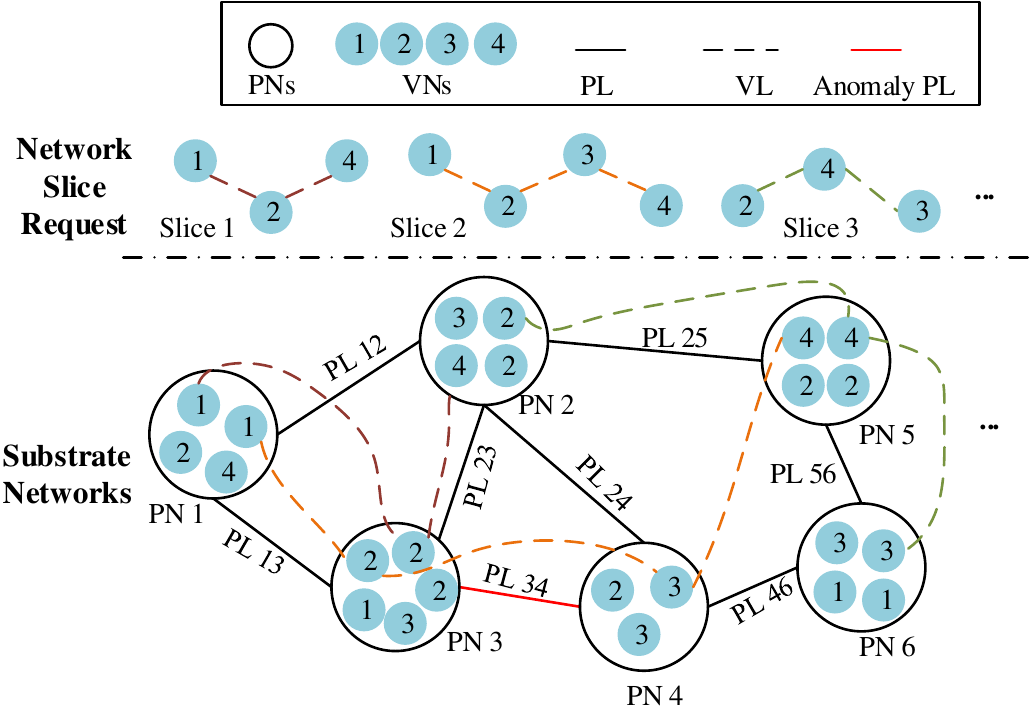}
\caption{The PL mapping schematic.}
\end{figure}

CCA is a widely used multivariate analysis algorithm \cite{jiang2017data}. Given two sets of random variables $\bm{U}$ and $\bm{Y}$ where

\begin{equation}
\begin{array}{l}
\bm{U} = \left[ {\begin{array}{*{20}c}
   {{\bm{u}}_1 }  \\
   {{\bm{u}}_2 }  \\
    \vdots   \\
   {{\bm{u}}_t }  \\
\end{array}} \right] = \left[ {\begin{array}{*{20}c}
   {u_{11} } & {u_{12} } &  \ldots  & {u_{1p} }  \\
   {u_{21} } & {u_{22} } &  \ldots  & {u_{2p} }  \\
    \vdots  &  \vdots  &  \ddots  &  \vdots   \\
   {u_{t1} } & {u_{t2} } &  \ldots  & {u_{tp} }  \\
\end{array}} \right]\\
\;\;\;\;\;\;\;\;\;\;\;\;\;\;\;\;\;\;\;\;\;\;\;\;\;\;\;\;\;\;\;\;\;\;\;\;\;(t\;{\rm{samples}}\;p\;{\rm{variables}}), \\
\bm{Y} = \left[ {\begin{array}{*{20}c}
   {{\bm{y}}_1 }  \\
   {{\bm{y}}_2 }  \\
    \vdots   \\
   {{\bm{y}}_t }  \\
\end{array}} \right] = \left[ {\begin{array}{*{20}c}
   {y_{11} } & {y_{12} } &  \ldots  & {y_{1d} }  \\
   {y_{21} } & {y_{22} } &  \ldots  & {y_{2d} }  \\
    \vdots  &  \vdots  &  \ddots  & {\rm{ \ldots }}  \\
   {y_{t1} } & {y_{t2} } &  \ldots  & {y_{td} }  \\
\end{array}} \right]\\
\;\;\;\;\;\;\;\;\;\;\;\;\;\;\;\;\;\;\;\;\;\;\;\;\;\;\;\;\;\;\;\;\;\;\;\;\;(t\;{\rm{samples}}\;d\;{\rm{variables}}), \\
 \end{array}
\end{equation}
CCA algorithm attempts to find canonical correlation vectors ${\bm{J}}$ and ${\bm{L}}$, which maximize the correlation $\rho$ between ${\bm{J}}^{\rm{T}}\bm{U}$ and ${\bm{L}}^{\rm{T}}\bm{Y}$, i.e.,
\begin{equation}
\begin{gathered}
  ({\bm{J}},{\bm{L}}) = \mathop {\arg \;\max }\limits_{({\bm{J}},{\bm{L}})} \;\rho _{({\bm{J}}^{\rm{T}} \bm{U})({\bm{L}}^{\rm{T}} \bm{Y})}  \hfill \\
  \;\;\;\;\;\;\;\;\; = \mathop {\arg \;\max }\limits_{({\bm{J}},{\bm{L}})} \;\frac{{{\bm{J}}^{\rm{T}} \bm{\Sigma} _{\bm{UY}} {\bm{L}}}}
{{({\bm{J}}^{\rm{T}} \bm{\Sigma} _{\bm{U}} {\bm{J}})^{1/2} ({\bm{L}}^{\rm{T}} \bm{\Sigma} _{\bm{Y}} {\bm{L}})^{1/2} }}, \hfill \\
\end{gathered}
\end{equation}
where $\bm{\Sigma} _{( \cdot )}$ denotes the covariance matrix. The solution to problem (22) can be obtained by implementing singular value decomposition on the matrix $\bm{K}$, which is shown as
\begin{equation}
{\bm{K}} = \bm{\Sigma} _{\bm{U}}^{ - 1/2}\bm{\Sigma} _{\bm{UY}} \bm{\Sigma} _{\bm{Y}}^{ - 1/2}  = {\bm{R}}\bm{\Sigma} {\bm{V}}^{\rm{T}},
\end{equation}
with $
\bm{R} = (\bm{r}_1 ,...,\bm{r}_p ),\;\bm{V} = (\bm{v}_1 ,...,\bm{v}_q ),\bm{\Sigma}  = \left[ {\begin{array}{*{20}c}
   {\bm{\Sigma} _\kappa  } & 0  \\
   0 & 0  \\
\end{array} } \right]$,
where $\bm{\Sigma} _\kappa = {\text{ diag}}(\rho _1 ,...,\rho _\kappa )$, $\kappa $ denotes the number of nonzero singular values and $\rho _i\;(i=1,...,\kappa)$ are canonical correlation coefficients. $\bm{r}_i\;(i=1,...,p)$ and $\bm{v}_i\;(i=1,...,q)$ are corresponding singular vectors.

The canonical correlation vectors can be derived as
\begin{equation}
{\bm{J}} = [{\bm{J}}_1 ,...,{\bm{J}}_p ] = \bm{\Sigma} _{\bm{U}}^{ - 1/2} {\bm{R}} \in \mathbb{R}^{p \times d},
\end{equation}
\begin{equation}
{\bm{L}} = [{\bm{L}}_1 ,...,{\bm{L}}_d ] = \bm{\Sigma} _{\bm{Y}}^{ - 1/2} {\bm{V}} \in \mathbb{R}^{d \times p}.
\end{equation}

Assume that PL $r$ and its both ends are corresponding to the physical path ${\text{PN}}_m \mathop  \to \limits^{{\text{PL}}_{m,l} } {\text{PN}}_{l}$. To detect the working state of PL $r$, i.e., ${\text{PL}}_{m,l}$ in a distributed manner, we need to find each SFC, whose virtual path ${\text{VN}}_m \mathop  \to \limits^{{\text{VL}}_{m,l} } {\text{VN}}_{l}$ is embedded to this physical path. Then, the correlation of measurements between ${\text{VN}}_m$ and ${\text{VN}}_{l}$ needs to be analyzed in its Network Slice Manager distributedly. Assume that the current measurements of ${\text{VN}}_m$ and ${\text{VN}}_{l}$ are represented by $\bm{u}$ and $\bm{y}$, respectively, then the key step for analyzing the correlation of measurements between neighbor VNs is to produce an anomaly detection residual, which is shown as follows:
\begin{equation}
{\bm{r}} = {\bm{J}}^{\rm{T}} \bm{u} - \bm{\Sigma} {\bm{L}}^{\rm{T}} \bm{y}.
\end{equation}

Then, the $T^2$ test for residual can be established as
\begin{equation}
T_r^2  = {\bm{r}}^{\rm{T}} \bm{\Sigma} _r^{ - 1} {\bm{r}},\;\;{\text{where}}\;\bm{\Sigma} _r = {\bm{I}}_p  - \bm{\Sigma}\bm{\Sigma} ^{\rm{T}}.
\end{equation}

Therefore, the correlation of measurements between neighbor VNs can be determined based on the following decision logic as
 \begin{equation}
\left\{ {\begin{array}{*{20}c}
   {T_r^2  \leqslant T_{r,{\text{cl}}}^2 \;\; \Rightarrow {\text{anomaly - free}}},  \hfill  \\
   {T_r^2  > T_{r,{\text{cl}}}^2 \; \Rightarrow {\text{anomaly}}},   \hfill \\
 \end{array} } \right.
\end{equation}
where $T_{r,{\text{cl}}}^2$ is the control limit of $T_r^2$, which represents the threshold that distinguishes between normal and anomalous correlation patterns.

\emph{Proposition 1}: Because the residual vector ${\bm{r}} = {\bm{J}}^{\rm{T}} {\bm{u}} - \bm{\Sigma} {\bm{L}}^{\rm{T}} {\bm{y}}$ has the minimum covariance, it is an optimal residual for detecting the variation of correlation between two sets of variables $\bm{u}$ and $\bm{y}$.

\emph{Proof}: See Appendix A.

Since new measurements will be generated in each period, to obtain the real-time canonical correlation vectors ${\bm{J}}(t)$ and ${\bm{L}}(t)$ for the online PL anomaly detection, we need to calculate covariances in formula (23) with all the historical data at each time $t$. Its computational complexity will become extremely high over time. To reduce the computational complexity and the storage consumption, we derive a new method to update covariances in formula (23), with which the Network Slice Manager only needs to store the covariance matrices and mean vectors of current VN data instead of all the historical data.

\emph{Proposition 2}: Assume that the covariance matrix of ${\bm{U}}(t)$ is $\bm{\Sigma} _{\bm{U}(t)}  = \frac{1}{{t - 1}}\left[ {\begin{array}{*{20}c}
   {a_{11} } & {a_{12} } &  \ldots  & {a_{1p} }  \\
   {a_{21} } & {a_{22} } &  \ldots  & {a_{2p} }  \\
    \vdots  &  \vdots  &  \ddots  &  \vdots   \\
   {a_{p1} } & {a_{p2} } &  \ldots  & {a_{pp} }  \\
\end{array}} \right]$  and the mean vector of  ${\bm{U}}(t)$ is $[c_1(t),c_2(t),...,c_p(t)]$ at time $t$, if the measurements of $\bm{u}$ is ${\bm{u}}_{t + 1}  = (u_{(t + 1)1},u_{(t + 1)2},...,u_{(t + 1)p} )$ at time $t+1$, the covariance matrix of ${\bm{U}}(t+1)$ can be computed by
\begin{equation}
\begin{gathered}
\bm{\Sigma} _{\bm{U}(t+1)}  =  \hfill  \\
 \frac{1}{t} \times \left[ {\begin{array}{*{20}c}
   {a_{11}  + m_{11} } & {a_{12}  + m_{12} } &  \ldots  & {a_{1p}  + m_{1p} }  \\
   {a_{21}  + m_{21} } & {a_{22}  + m_{22} } &  \ldots  & {a_{2p}  + m_{2p} }  \\
    \vdots  &  \vdots  &  \ddots  &  \vdots   \\
   {a_{p1}  + m_{p1} } & {a_{p2}  + m_{p2} } &  \ldots  & {a_{pp}  + m_{pp} }  \\
\end{array}} \right], \\
\end{gathered}
\end{equation}
where $m_{i,j}  = \frac{{t(c_i (t) - u_{(t + 1)i} )(c_j (t) - u_{(t + 1)j} )}}{{t + 1}}(1 \le i,j \le p)$.

\emph{Proof}: See Appendix B.

Similarly, assume that the covariance matrix between $\bm{U}(t)$ and $\bm{Y}(t)$ is $\bm{\Sigma} _{\bm{U}(t)\bm{Y}(t)}  = \frac{1}{{t - 1}}\left[ {\begin{array}{*{20}c}
   {e_{11} } & {e_{12} } &  \ldots  & {e_{1d} }  \\
   {e_{21} } & {e_{22} } &  \ldots  & {e_{2d} }  \\
    \vdots  &  \vdots  &  \ddots  &  \vdots   \\
   {e_{p1} } & {e_{p2} } &  \ldots  & {e_{pd} }  \\
\end{array}} \right]$  and the mean vector of $\bm{Y}(t)$ is $[f_1 (t),f_2 (t),...,f_d (t)]$ at time $t$, if the measurements of ${\bm{y}}$ is ${\bm{y}}_{t + 1}  = (y_{(t + 1)1},y_{(t + 1)2},...,y_{(t + 1)d} )$ at time $t+1$, the covariance matrix between $\bm{U}(t+1)$ and $\bm{Y}(t+1)$ can be computed by
\begin{equation}
\begin{gathered}
\bm{\Sigma} _{\bm{U}(t+1)\bm{Y}(t+1)}   =  \hfill  \\
 \frac{1}{t} \times \left[ {\begin{array}{*{20}c}
   {e_{11}  + n_{11} } & {e_{12}  + n_{12} } &  \ldots  & {e_{1d}  + n_{1d} }  \\
   {e_{21}  + n_{21} } & {e_{22}  + m_{22} } &  \ldots  & {e_{2d}  + n_{2d} }  \\
    \vdots  &  \vdots  &  \ddots  &  \vdots   \\
   {e_{p1}  + n_{p1} } & {e_{p2}  + n_{p2} } &  \ldots  & {e_{pd}  + n_{pd} }  \\
\end{array}} \right], \\
 \end{gathered}
\end{equation}
where $n_{ij}  = \frac{{t(c_i (t) - u_{(t + 1)i} )(d_j (t) - y_{(t + 1)j} )}}{{t + 1}}(1 \le i,(j) \le p,(d))$.

Therefore, the covariance matrices at time $t+1$ can be calculated by only keeping the covariance matrices and mean vectors at time $t$. Then, the canonical correlation vectors ${\bm{J}}(t+1)$ and ${\bm{L}}(t+1)$ can be derived to produce the optimal residual without storing all the historical data, so large storage and computing resources can be saved.

The detailed steps of the CCA-based distributed online PL anomaly detection algorithm are shown in Algorithm 2.

\begin{algorithm}[t]
\caption{CCA-based distributed online PL anomaly detection algorithm}
\begin{algorithmic}[1]
\REQUIRE Initial number of labeled samples $t$, measurements $\bm{U}(t)$ and $\bm{Y}(t)$ of ${\text{VN}}_m$ and ${\text{VN}}_{l}$ for each ${\text{VN}}_m \mathop  \to \limits^{{\text{VL}}_{m,l} } {\text{VN}}_{l}$ mapped to ${\text{PN}}_m \mathop  \to \limits^{{\text{PL}}_{m,l} } {\text{PN}}_{m,l}$, control limit $T_{r,{\text{cl}}}^2$ and number of iterations $T$
\STATE  Compute the covariance matrices and mean vectors of $\bm{U}(t)$ and $\bm{Y}(t)$: $\bm{\Sigma} _{\bm{U}(t)}$, $\bm{\Sigma} _{\bm{Y}(t)}$, $\bm{\Sigma} _{\bm{U}(t)\bm{Y}(t)}$,  $[c_1(t),c_2(t),...,c_p(t)]$ and $[f_1 (t),f_2 (t),...,f_d (t)]$
\FOR {$t = t+1:T$}
\FOR {each ${\text{VN}}_l \mathop  \to \limits^{{\text{VL}}_{l,l + 1} } {\text{VN}}_{l + 1}$}
\STATE Compute $\bm{\Sigma} _{\bm{U}(t)}$, $\bm{\Sigma} _{\bm{Y}(t)}$ and $\bm{\Sigma} _{\bm{U}(t)\bm{Y}(t)}$ according to formulas (29) and (30)
\STATE Implement singular value decomposition on the matrix $\bm{K}(t)$ according to formula (23)
\STATE Compute canonical correlation vectors $\bm{J}(t)$ and $\bm{L}(t)$ according to formulas (24) and (25)
\STATE Produce the optimal anomaly detection residual ${\bm{r}}(t)$ according to formula (26)
\STATE Establish the $T^2$ test: $T_{r(t)}^2  = {\bm{r}}(t)^{\rm{T}} \bm{\Sigma} _{r(t)}^{ - 1} {\bm{r}}(t)$
\ENDFOR
\IF {$T_{r(t)}^2  \leqslant T_{r,{\text{cl}}}^2$ for each ${\text{VN}}_m \mathop  \to \limits^{{\text{VL}}_{m,l} } {\text{VN}}_{l}$}
\STATE Determine that ${\text{PL}}_{m,l}$ (or PL $r$) is normal, and update covariance matrices and mean vectors
\ELSE
\STATE Determine that ${\text{PL}}_{m,l}$ (or PL $r$) is anomalous, and reserve covariance matrices and mean vectors at time $t-1$ and discard current ones
\ENDIF
\ENDFOR
\end{algorithmic}
\end{algorithm}

 We usually deal with the anomaly detection as a special classification problem by assuming that the entire dataset contains only normal samples. When the number of anomalous data contained in training samples increases, the performance of classical CCA method will decline greatly. By comparison, the proposed algorithm is more robust for it can eliminate the impact of anomalous data on the detection accuracy in each iteration (step 13).

 The computation of CCA algorithm mainly focuses on the covariance calculation. When the number of samples reaches $T$ and the number of VLs mapped to PL $r$ is $|J_r|$, the time and storage complexity of the classical CCA algorithm are $O(T|J_r|(p^2  + pd + d^2 ))$ and $O(T|J_r|(p + d + 2))$, respectively. For the distributed online mode, the time and storage complexity of the proposed algorithm are $O(p^2  + pd + d^2 )$ and $O(p + d)$, respectively.

\section{Simulation Results}
\subsection{Evaluation Metrics}

In this section, numerical simulations on both synthetic and real-world network datasets are executed to evaluate the proposed distributed online PN and PL anomaly detection algorithms. The performance analysis of the proposed algorithms is evaluated using various metrics such as precision, recall and f1-socre, which can be formulated by confusion matrix as presented in Table I.
\begin{table}[htbp]
\centering
\caption{Confusion matrix}
\renewcommand\arraystretch{1.5}
\scalebox{1}{\begin{tabular}{c|c|c|c}
\hline
\hline
\multicolumn{2}{c|}{\multirow{2}*{~}}&\multicolumn{2}{c}{Predicted result}\\
\cline{3-4}
\multicolumn{2}{c|}{~}&normal&anomalous\\
\hline
\multirow{2}*{Actual result}&normal&True Negative (TN)&False Positive (FP)\\
\cline{2-4}
&anomalous&False Negative (FN)&True Positive (TP)\\
\hline
\end{tabular}}
\end{table}

TN, FN, FP and TP are defined as follows: TN represents the number of normal working states which are identified correctly. FN indicates the number of anomalies which are not identified. FP summarizes the normal working states that have been judged as anomalies. TP represents the number of anomalies which are identified correctly.
\begin{itemize}
\item \emph{Accuracy} is the proportion of correctly predicted PN or PL working states to the total ones.
\begin{equation}
Accuracy = \frac{{TP + TN}}{{TP + TN + FP + FN}}
\end{equation}

\item \emph{Precision} represents the proportion of correctly predicted anomalous PN or PL working states to the total predicted anomalous ones.
\begin{equation}
Precision = \frac{{TP}}{{TP + FP}}
\end{equation}

\item \emph{Recall} represents the proportion of correctly predicted anomalous PN or PL working states to the total actual anomalous ones.
\begin{equation}
Recall = \frac{{TP}}{{TP + FN}}
\end{equation}

\item \emph{F1-score} is the weighted average of the precision and recall metrics.
\begin{equation}
F1-score = 2 \times \frac{{precison \times recall}}{{precison + recall}}
\end{equation}

\end{itemize}

Accuracy is an effective evaluation metric when the datasets are balanced, but for imbalanced ones it may give biased evaluation results. Compared to accuracy, precision and recall are less biased metrics for the evaluation of imbalanced datasets. Therefore, precision, recall and f1-score metrics are mainly used to analyze the performance of the proposed anomaly detection algorithms.

\subsection{Synthetic Dataset}

To generate synthetic data to validate the effectiveness of the proposed PN and PL anomaly detection algorithms, a simple virtualized network slicing scenario with 10 PNs and 6 SFCs has been established in MATLAB platform. Each SFC is assumed to contain 4 $\sim$ 6 VNs \cite{Fu2019service}. Three different service requests are contained to simulate the diverse service types in network slicing. VNs and VLs in SFCs are randomly mapped to the substrate network to provide end-to-end services for different user types. When PNs and PLs are in normal states, their processing capacity is in normal range. To simulate the anomalous cases, the loss rates of processing capacity for PNs and PLs, which follows the Gaussian distribution $N(\mu ,\sigma ^2 )$ with the mean $\mu  = 0.5$ and the variance $\sigma ^2  = 0.01$, are randomly injected into the networks. Then, the simulated data can be acquired in each period, including processing rate, data flow, queuing delay, processing delay, etc. Table II presents the main parameters used for simulation.
\begin{table}[t]
\centering
\caption{Simulation parameters}
\scalebox{1.05}{\begin{tabular}{c c}
\hline
\hline
 Parameter & Value \\
\hline
 PN number   & 10 \\
 SFC number  & 6 \\
 VN number in each SFC   &  4$\sim$6 \\
Service 1 (arrive rate, packet size)    &   (10packets/s, 200kbit/packets) \\
Service 2 (arrive rate, packet size)  &  (100packets/s, 10kbit/packets) \\
Service 3 (arrive rate, packet size)    &   (500packets/s, 1kbit/packets) \\
Random feature space   &   100-dimensional\\
\hline
\end{tabular}}
\end{table}

\begin{figure}[t]
    \centering
    \subfigure[Precision]{\label{Fig:R1}
    \includegraphics[width=2.9in]{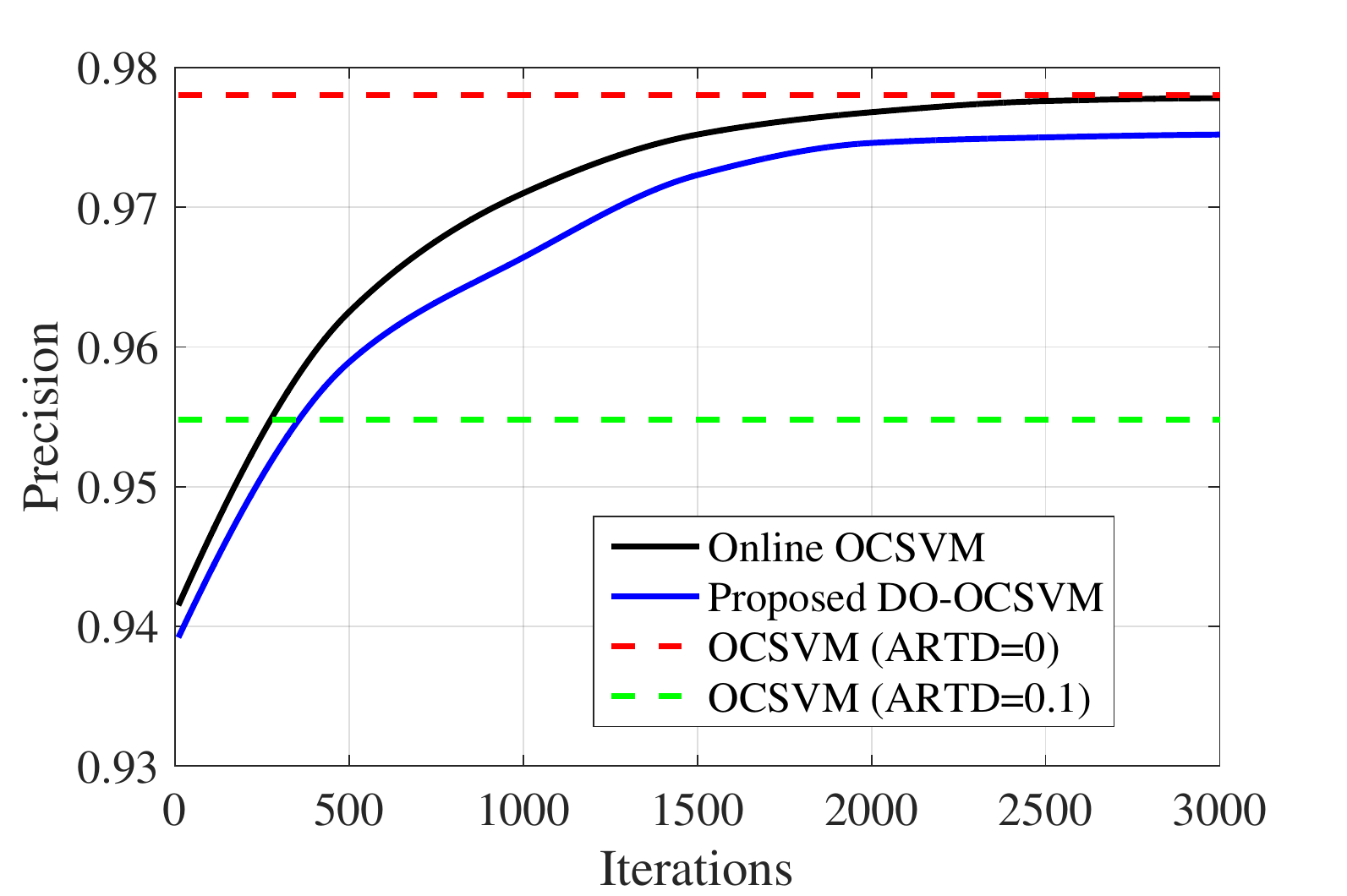}}
    \subfigure[Recall]{\label{Fig:R2}
    \includegraphics[width=2.9in]{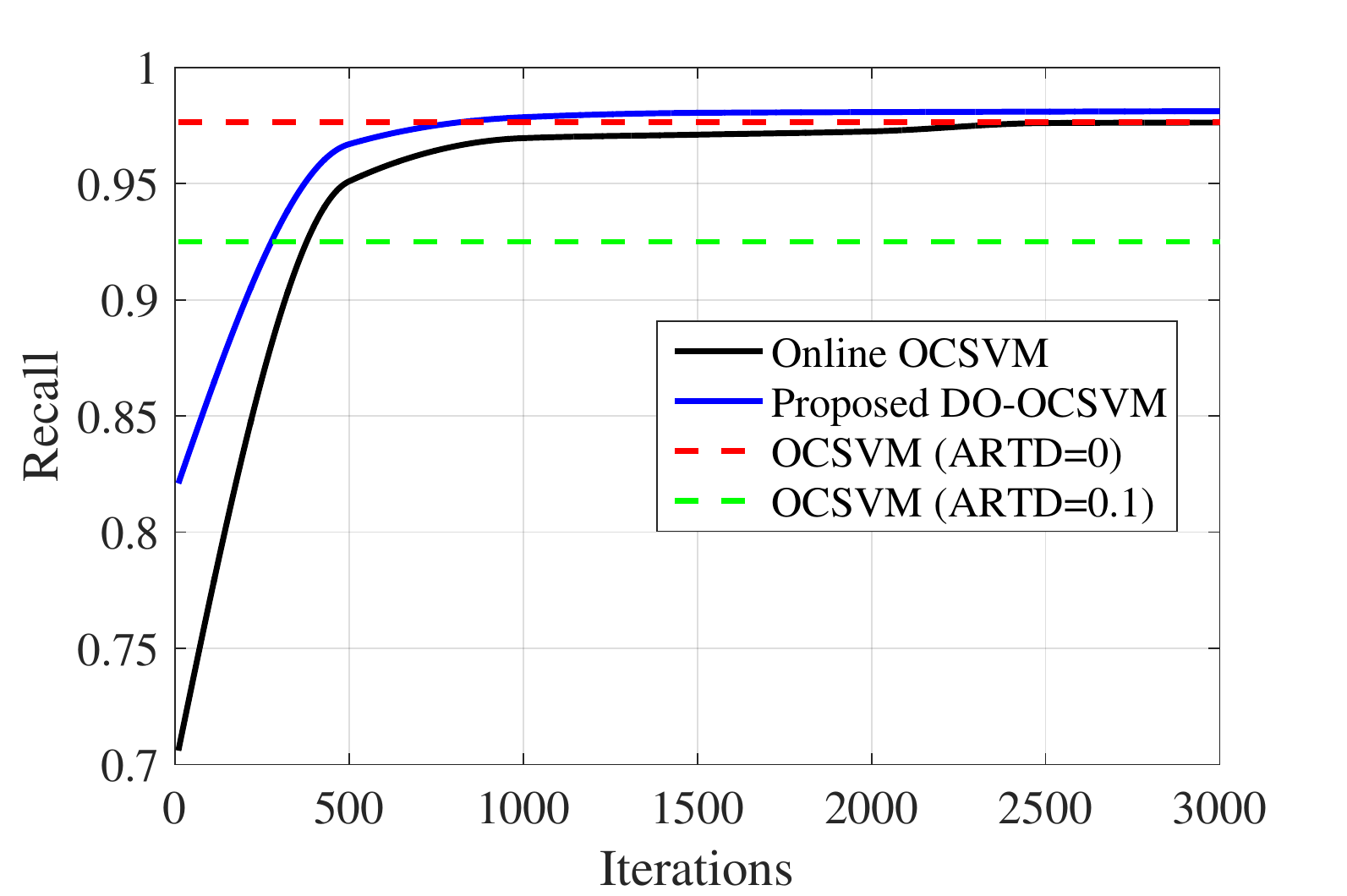}}\\
 \subfigure[F1-score]{\label{Fig:R3}
  \includegraphics[width=2.9in]{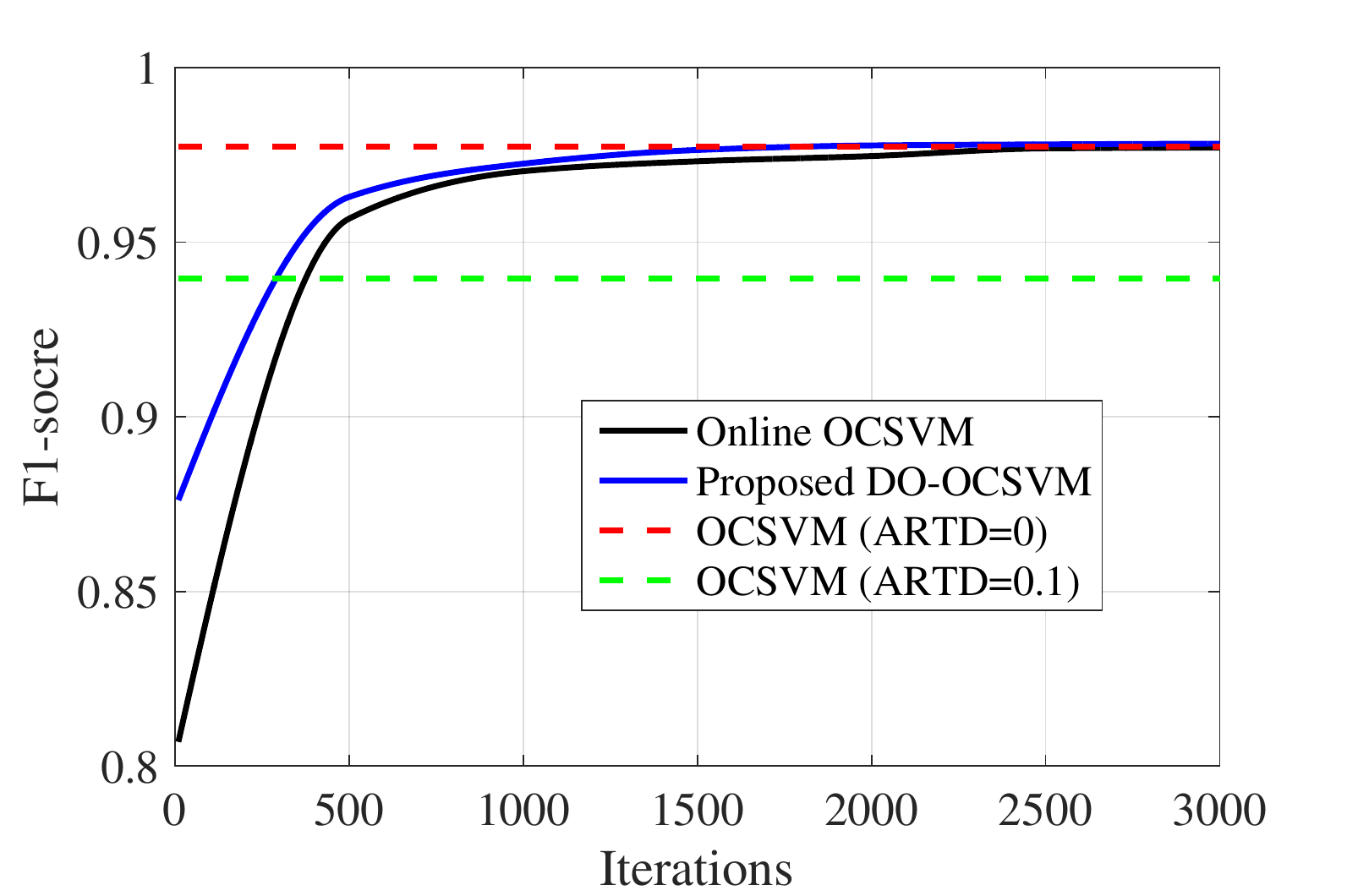}}\\
    \caption{Performance comparison among different PN anomaly detection algorithms on synthetic dataset.}
\end{figure}

For comparison, we have simulated the OCSVM based \cite{jiang2013anomaly} and online OCSVM based \cite{gomez2011adaptive} anomaly detection methods. Fig. 3 shows the comparison results of precision, recall and f1-score among the OCSVM, online OCSVM and the proposed distributed online OCSVM (DO-OCSVM) algorithms. The presented results are averaged total 500 Monte Carlo runs. To verify the impact of anomalous data contained in training samples on the classical OCSVM algorithm, we have introduced the concept of anomaly ratio in training data (ARTD). From Fig. 3, compared with the classical OCSVM and the online OCSVM algorithms, the proposed DO-OCSVM has a higher recall and f1-score, but its precision is a little lower. This is because in the proposed DO-OCSVM algorithm, as long as measurements of one VN mapped to a PN are detected as anomaly, this PN will be determined to be anomalous. Therefore, the proposed algorithm can detect the anomalous PNs more accurately, but the possibility of wrongly determining normal PNs as anomalous ones also increases. Since we pay more attention to the recall in anomaly detection, a slight drop in precision is acceptable. Note that when the training samples contain anomalous data (ARTD=0.1), the performance of the classical OCSVM in three metrics will be  greatly degraded. However, it is costly to obtain large amounts of accurate labeled data in actual networks. Fortunately, the proposed DO-OCSVM algorithm can eliminate the impact of anomalous data on the detection accuracy during the iteration process without any labeled data, which makes it more economic and robust anomaly detection method.

Fig. 4 shows the convergence process of $\bm{w}$ and $\rho$ in the proposed DO-OCSVM algorithm. Since $\bm{w}$ is a 100-D vector, it is difficult to depict all the components. Therefore, we randomly select 5 components for illustration. From Fig. 4, we can see that $\bm{w}$ and $\rho$ in DO-OCSVM algorithm nearly remain stable after 1000 iterations. Note that the convergence process of parameters presented in Fig. 4 is basically consistent with that of precision, recall and f1-score shown in Fig. 3.

\begin{figure}[t]
 \centering
 \includegraphics[width=2.9in]{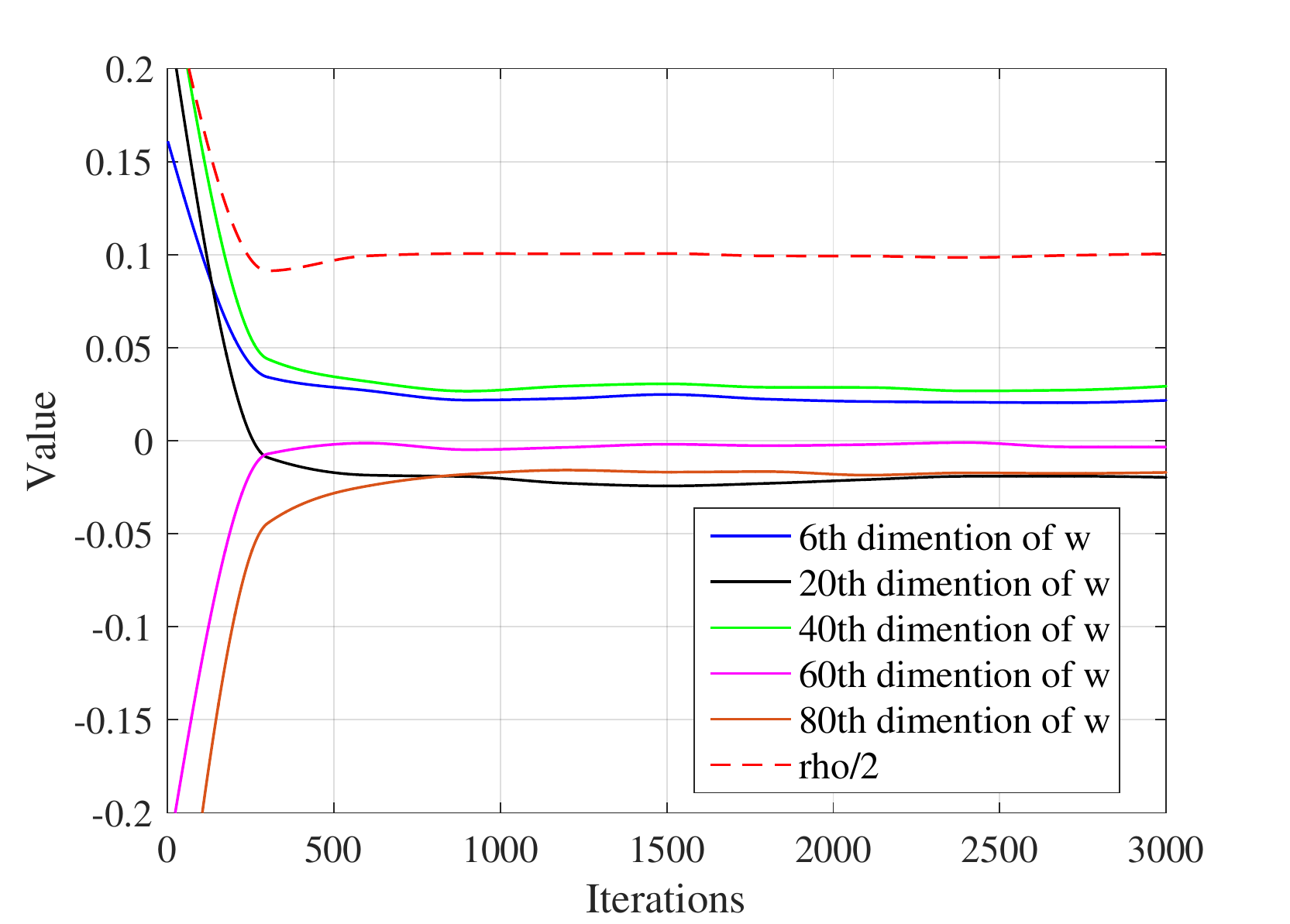}
\caption{Convergence process of $\bm{w}$ and $\rho$ in DO-OCSVM algorithm.}
\end{figure}

The performance comparison results between the CCA algorithm in \cite{chen2017fault}, online CCA algorithm and the proposed distributed online CCA (DO-CCA) algorithm for PL anomaly detection are shown in Fig. 5. We set the initial number of labeled samples $t=10$ and control limit $T_{r,{\rm{cl}}}^2  = 1$. The presented results are averaged total 500 Monte Carlo runs. To verify the impact of anomalous data in training samples on the classical CCA algorithm, we simulate the condition when ARTD=0.1. Observing Fig. 5, the proposed DO-CCA has a higher recall compared to the CCA and online CCA algorithms, which suggests that proposed DO-CCA algorithm can predict the anomalous PLs more accurately. Meanwhile, the precision of the proposed DO-CCA algorithm is a little lower, implying that the possibility of wrongly determined normal PLs as anomalous ones also has a little increase. Since we concentrate more on the performance of recall in anomaly detection, a slightly small precision is acceptable. Note that when the training samples contain anomalous data, the performance of the classical CCA algorithm will decline greatly. Since the proposed DO-CCA algorithm can eliminate the impact of anomalous data during the iteration process, it is more adaptive.
\begin{figure}[t]
    \centering
    \subfigure[Precision]{\label{Fig:R1}
    \includegraphics[width=2.9in]{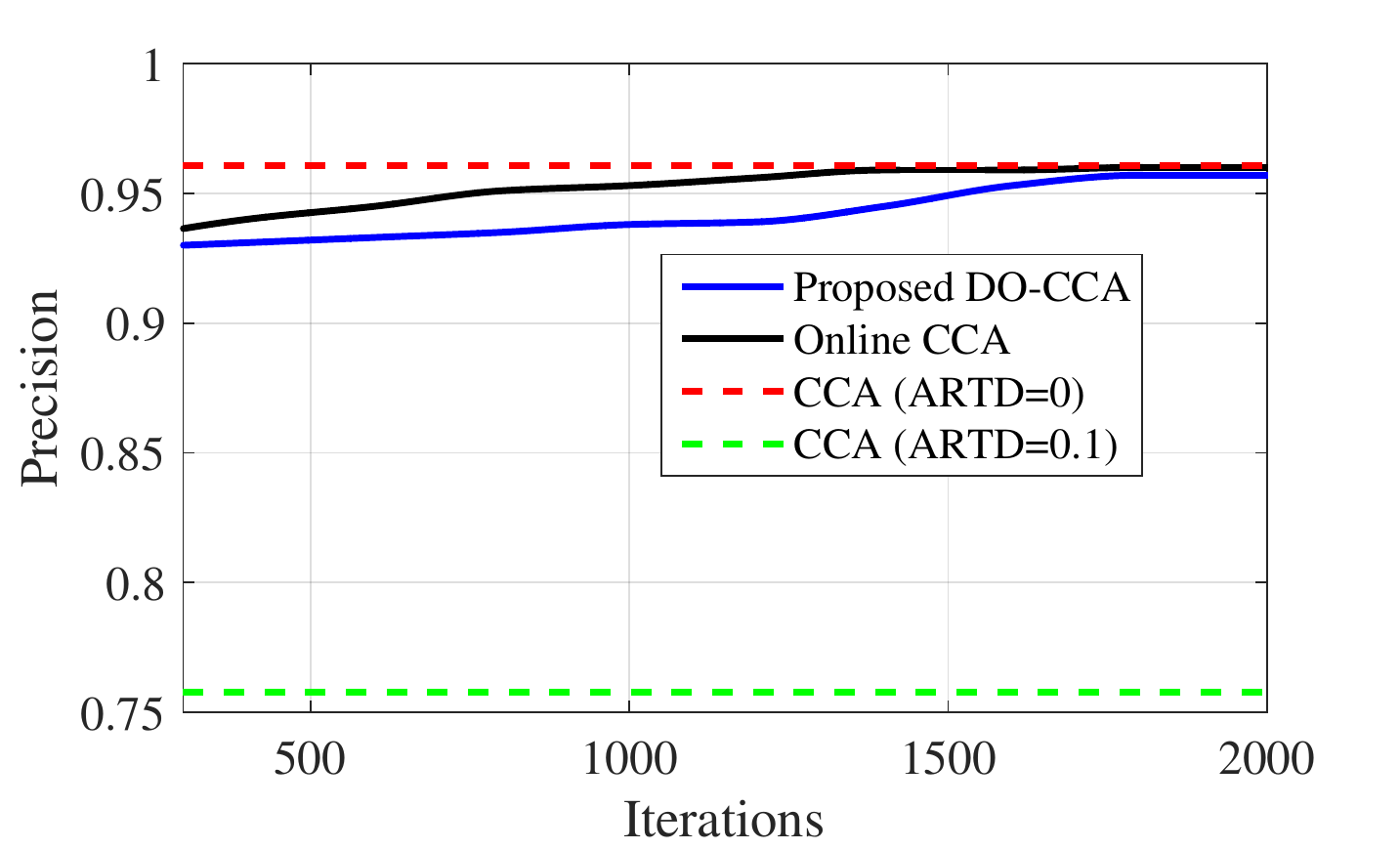}}
    \subfigure[Recall]{\label{Fig:R2}
    \includegraphics[width=2.9in]{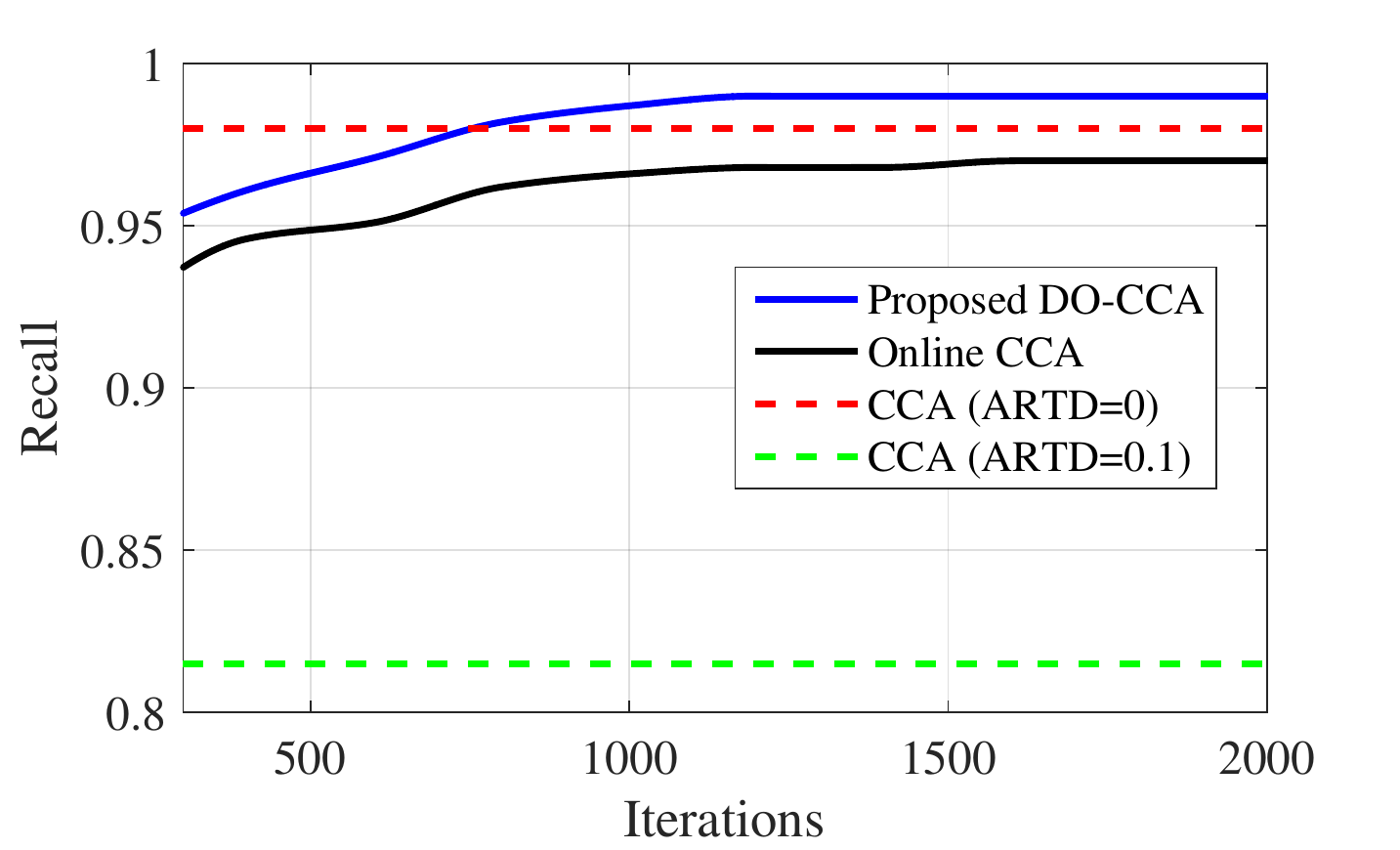}}\\
 \subfigure[F1-score]{\label{Fig:R3}
  \includegraphics[width=2.9in]{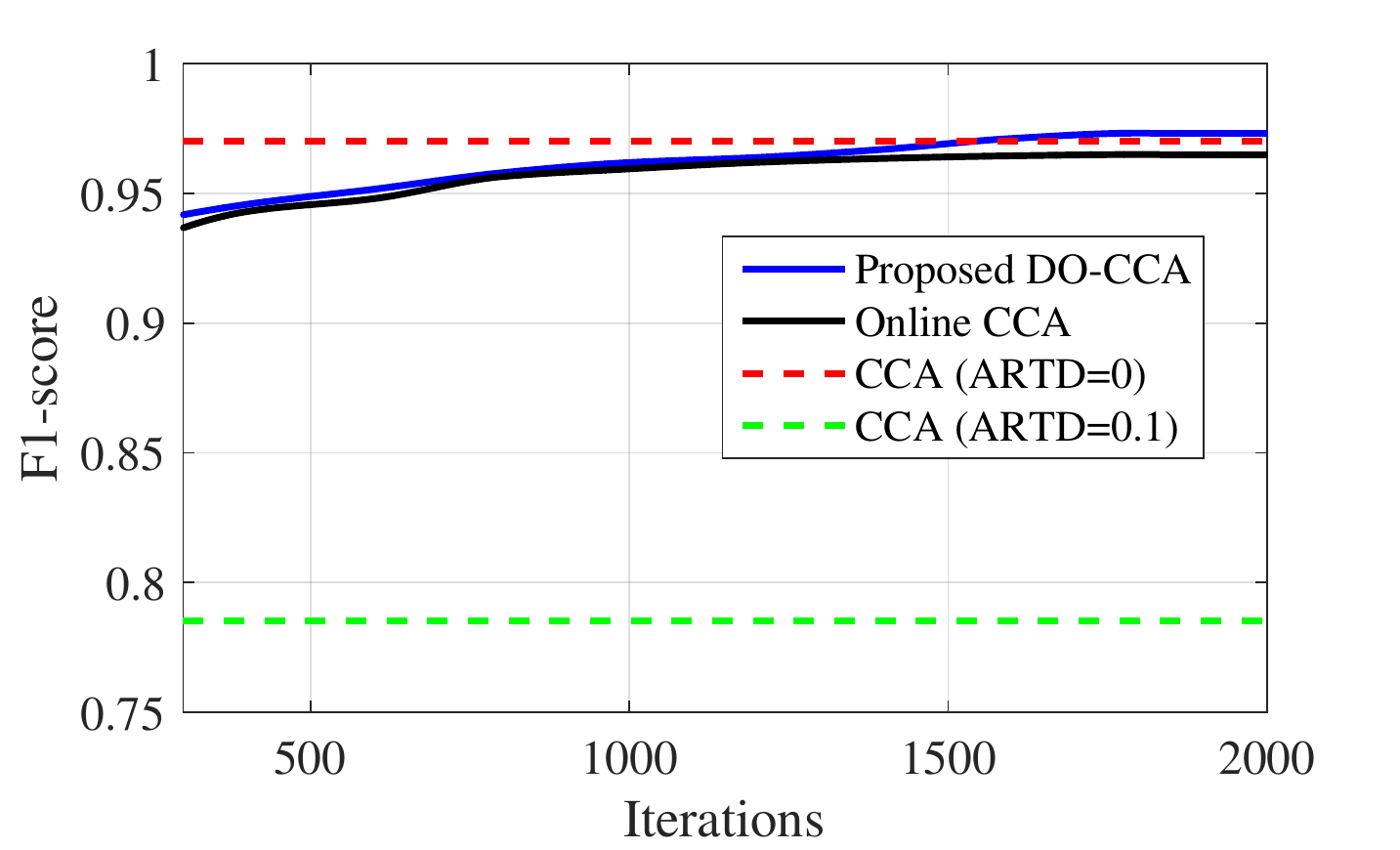}}\\
    \caption{Performance comparison among different PL anomaly detection algorithms on synthetic dataset.}
\end{figure}

\subsection{Real-World Network Dataset}
Except validating the proposed distributed online anomaly detection algorithms on the synthetic dataset from the simulated virtualized network slicing scenario, we further evaluate their performance on the real-world network dataset. The data rate performance measurements of NFV cloud native scaling for a media application \cite{2e1c-rd87-20}, which is selected from IEEEDataPort, include CPU, memory, disk and network metrics. The system monitoring period, i.e., the data sampling period was set to 15s. To evaluate the performance of different PN anomaly detection algorithms, we introduce four cases to simulate the anomalies in PNs, including endless loop in CPU, memory leak, Disk I/O fault and network congestion.

For the purpose of further comparison, we also evaluate the performance of two state-of-the-art classical anomaly detection methods on the four anomaly cases, which are \emph{k}-nearest neighbor (\emph{k}-NN) method \cite{Sridhar2000efficient} and local outlier factor (LOF) \cite{Markus2000lof} algorithm. The comparison results of different anomaly detection algorithms on four anomaly cases are summarized in Table III. The results show that the online OCSVM and DO-OCSVM algorithms have comparable or better detection performance than \emph{k}-NN and LOF. For example, under the case of Disk I/O fault, the detection performance of LOF is significantly lower than that of the other three algorithms. The reason is that anomalous disk read or write rate is not much different from the normal one, so the local density of monitored measurements has only a little fluctuation, which is difficult to capture for the LOF algorithm. Besides, LOF has a better performance on detecting endless loop in CPU, which indicates that LOF is more sensitive to the fluctuant metrics. By contrast, our proposed DO-OCSVM based PN anomaly detetion algorithm has the best detection performance on the four anomaly cases.

\begin{table*}[htbp]
\centering
\caption{The comparison results of different anomaly detection algorithms on four anomaly cases}
\renewcommand\arraystretch{1.3}
\scalebox{0.8}{\begin{tabular}{c| c c c| c c c| c c c| c c c}
\hline
\hline
\multirow{2}*{Anomalies}&\multicolumn{3}{c|}{\emph{k}-NN}&\multicolumn{3}{|c|}{LOF}&\multicolumn{3}{c}{Online OCSVM}&\multicolumn{3}{|c}{Proposed DO-OCSVM}\\
\cline{2-13}
&Recall&Precision&F1-score&Recall&Precision&F1-score&Recall&Precision&F1-score&Recall&Precision&F1-score\\
\hline
Endless loop in CPU&0.951&0.975&0.963&0.905&0.960&0.932&0.984&0.961&0.972&1.00&1.00&1.00\\
\hline
Memory Leak&0.903&0.939&0.923&0.894&0.849&0.871&0.976&0.934&0.955&0.988&0.969&0.978\\
\hline
Disk I/O fault&0.979&0.977&0.976&0.588&0.600&0.594&0.976&0.931&0.953&0.979&0.960&0.969\\
\hline
Network congestion&0.858&0.965&0.908&0.808&0.727&0.767&1.00&0.970&0.986&1.00&0.977&0.988\\
\hline
\end{tabular}}
\end{table*}

 \begin{figure}[t]
 \centering
\includegraphics[width=2.9in]{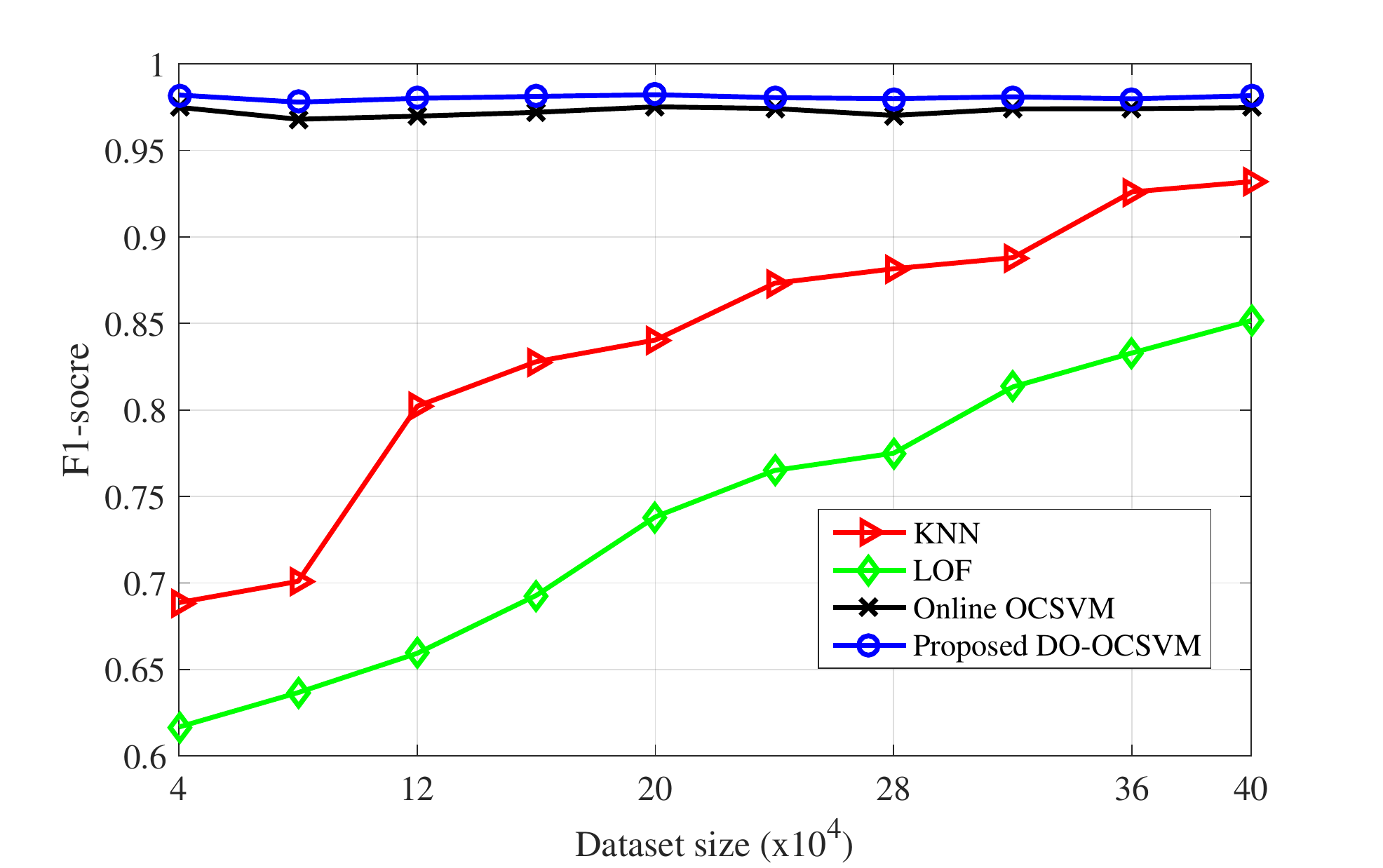}
\caption{F1-score comparison among different PN anomaly detection algorithms versus different sizes of real-world network dataset.}
\end{figure}

 \begin{figure}[t]
 \centering
\includegraphics[width=2.9in]{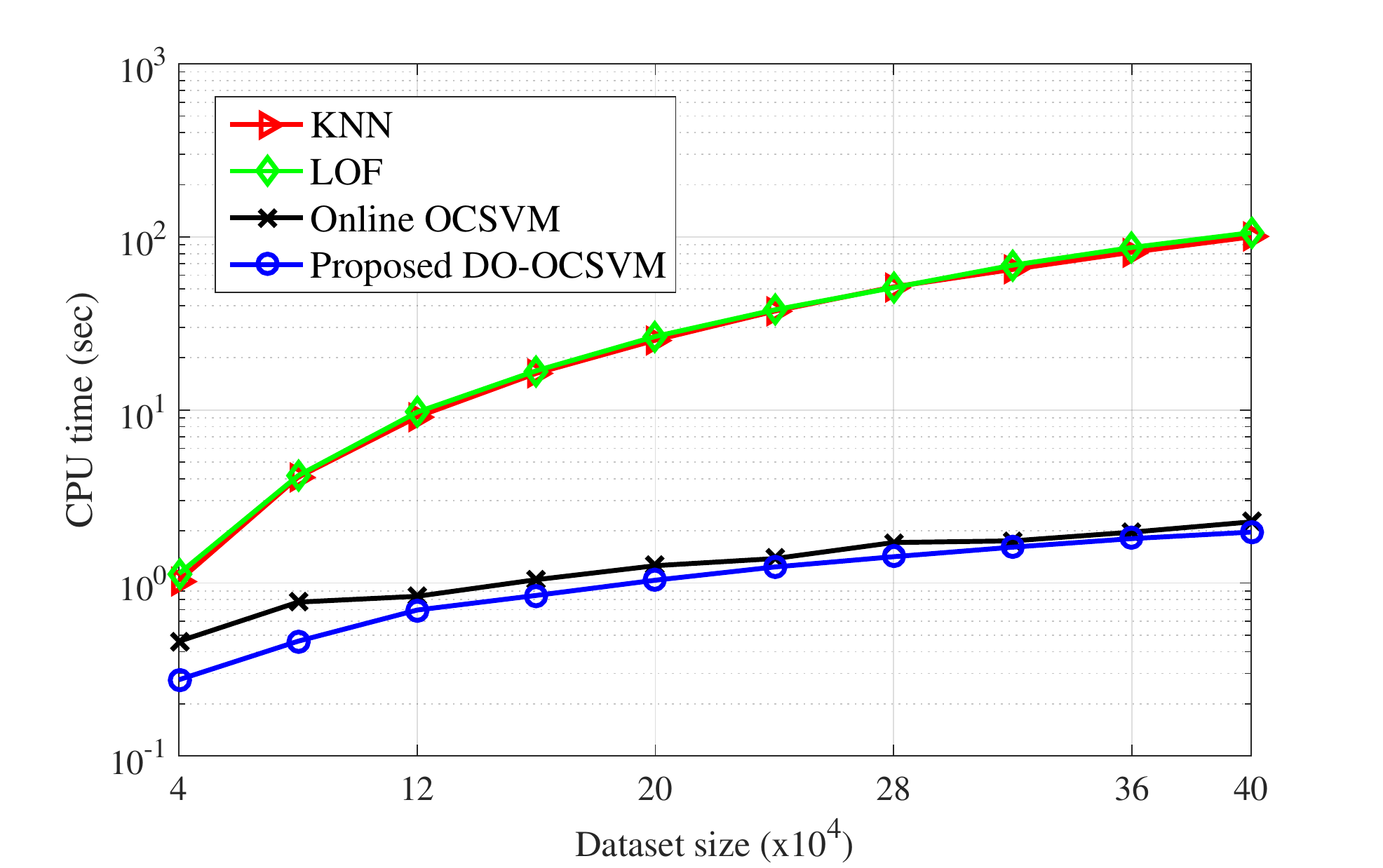}
\caption{CPU time comparison among different PN anomaly detection algorithms versus different sizes of real-world network dataset.}
\end{figure}

 \begin{figure}[t]
 \centering
\includegraphics[width=2.9in]{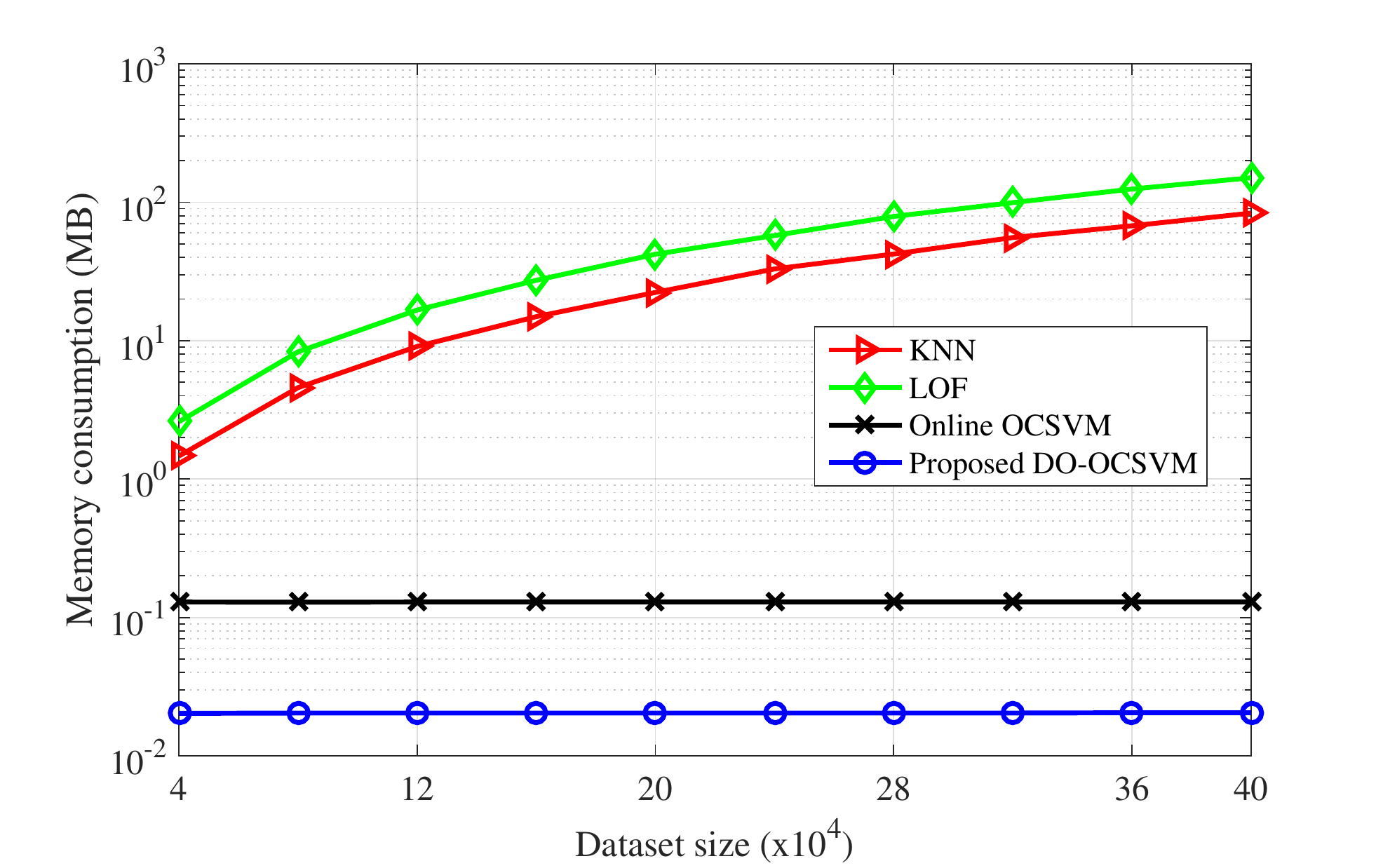}
\caption{Memory consumption comparison among different PN anomaly detection algorithms versus different sizes of real-world network dataset.}
\end{figure}

 \begin{figure}[t]
 \centering
\includegraphics[width=2.9in]{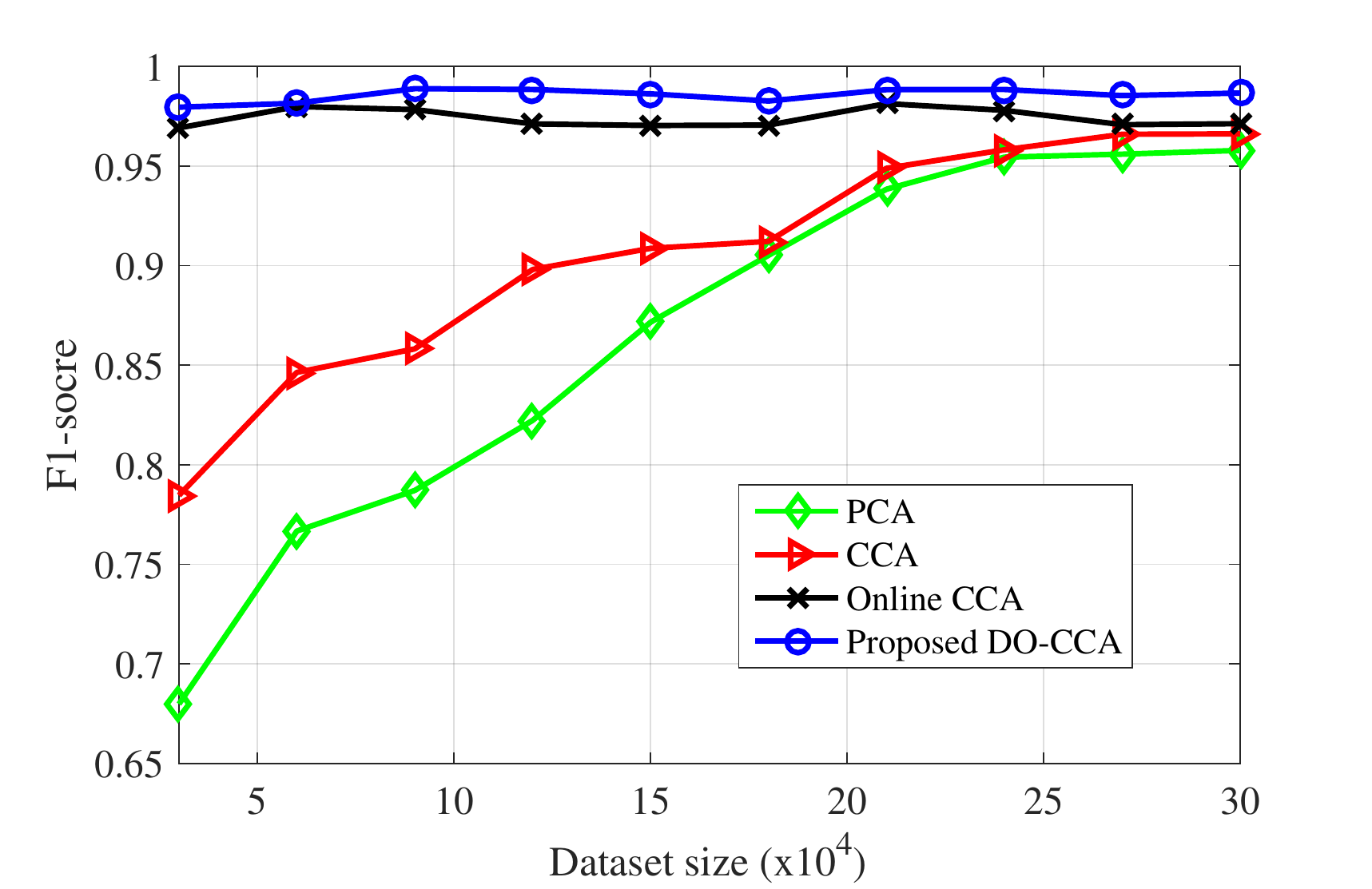}
\caption{F1-score comparison among different PL anomaly detection algorithms versus different sizes of real-world network dataset.}
\end{figure}

The f1-score, which is the weighted average of the precision and recall metrics, is used to evaluate the overall performance of different anomaly detection algorithms with different datatset sizes. The corresponding results are shown in Fig. 6. Furthermore, we also record the CPU time and memory consumption of each detection method with different dataset sizes, and the results are shown in Figs. 7 and 8, respectively. It should be noted that for the proposed distributed detection scheme, we only take the average CPU time and memory consumption of every distributed manager into account. Observing the simulation results depicted in Figs. 6-8, it is obvious that the online OCSVM and DO-OCSVM algorithms take much less CPU time and memory consumption than the classical \emph{k}-NN and LOF methods, and obtain better detection performance. The f1-score of \emph{k}-NN and LOF methods increases as the dataset size grows. By contrast, the online OCSVM and DO-OCSVM algorithms are not sensitive to the dataset sizes. Besides, the required CPU time and memory consumption of \emph{k}-NN and LOF methods increase exponentially as the dataset size grows. For the online OCSVM and DO-OCSVM algorithms, the required memory consumption remains nearly unchanged, and the required CPU time grows linearly due to the online detection mode.

For validating the performance of the proposed DO-CCA based PL anomaly detection algorithm on the real-world network dataset, in addition to the CCA and online CCA algorithms, we also evaluate the performance of the principal component analysis method (PCA)\cite{Steven2010on}, which is the state-of-the-art classical multivariate analysis method. The f1-score, the required CPU time and memory consumption are used to evaluate the overall performance of different PL anomaly detection algorithms. The corresponding results are shown in Figs. 9-11, respectively. Observing the simulation results depicted in Figs. 9-11, we can see that the online CCA and DO-CCA based PL anomaly detection algorithms take much less CPU time and memory consumption than the classical PCA and CCA methods, and obtain better detection performance. The f1-score of PCA and CCA methods increases, and the required CPU time and memory consumption increase linearly as the dataset size grows. By contrast, for the online CCA and DO-CCA algorithms, the f1-score, required CPU time and memory consumption all remain nearly unchanged as the dataset size grows. Besides, our proposed DO-CCA based PL anomaly detetion algorithm has the best detection performance with the lowest CPU time and memory consumption for the distributed online mode.

 \begin{figure}[t]
 \centering
\includegraphics[width=2.9in]{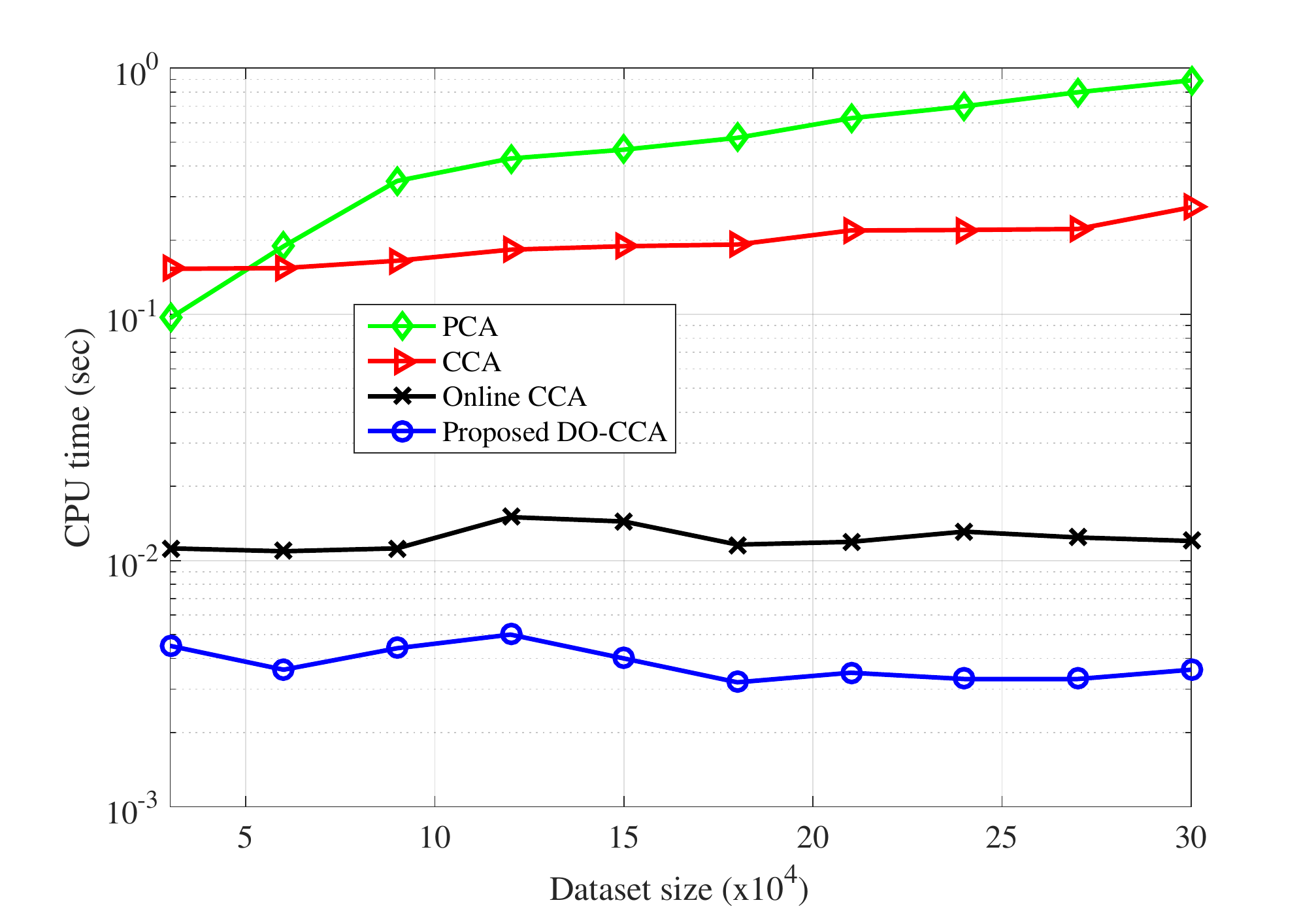}
\caption{CPU time comparison among different PL anomaly detection algorithms versus different sizes of real-world network dataset.}
\end{figure}

 \begin{figure}[t]
 \centering
\includegraphics[width=2.9in]{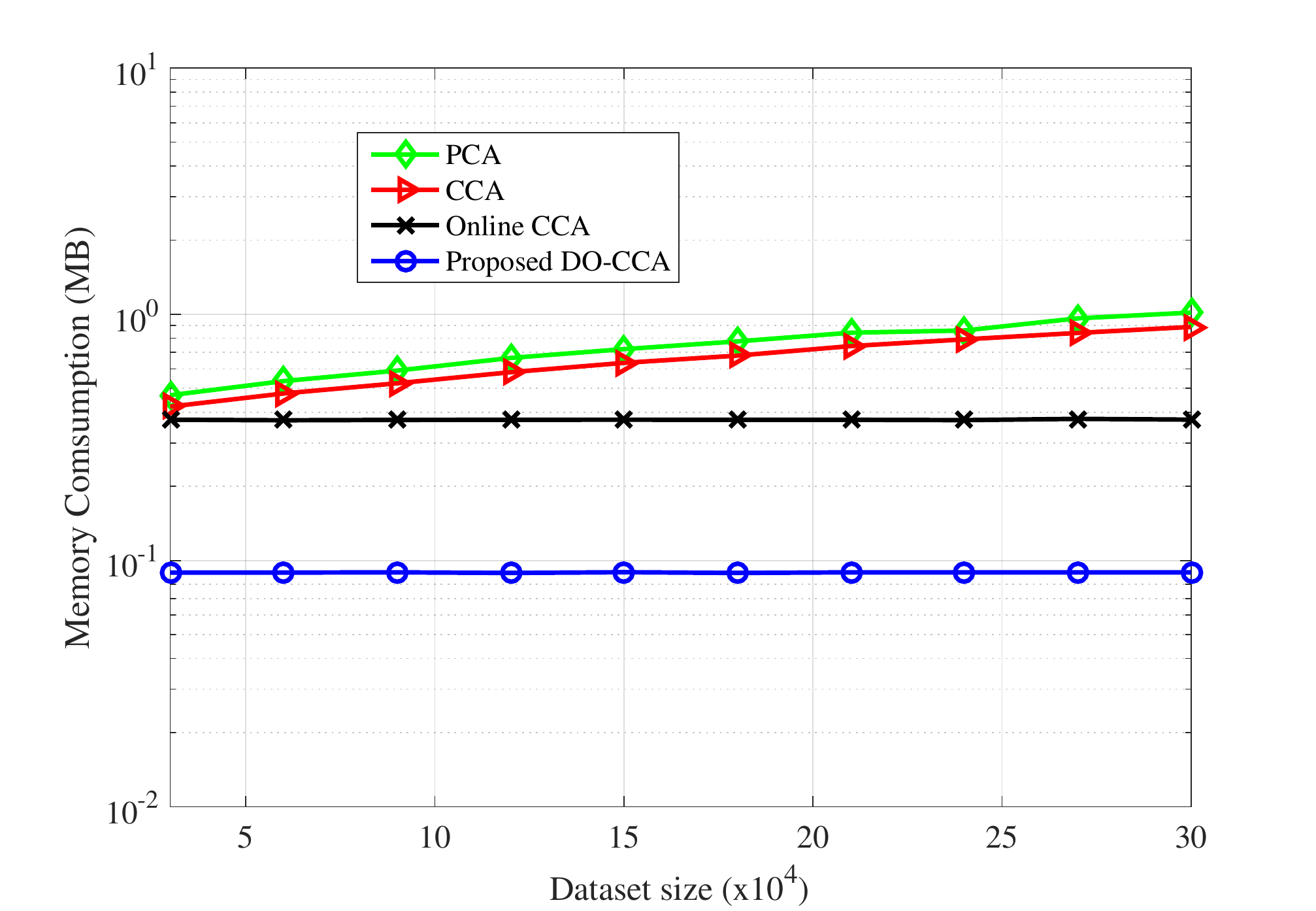}
\caption{Memory consumption comparison among different PL anomaly detection algorithms versus different sizes of real-world network dataset.}
\end{figure}

\section{Conclusions}
The accurate and rapid anomaly detection for PNs and PLs in substrate networks is the prerequisite for ensuring the performance of virtualized network slices. To realize the real-time anomaly detection for the substrate networks with low communication and storage cost, a distributed online PN anomaly detection algorithm was first proposed based on a decentralized OCSVM. This detection algorithm could identify the working state of a PN through analyzing the real-time measurements of each VN mapped to it in a distributed manner. Then, we proposed a CCA-based distributed online algorithm to realize the PL anomaly detection. This algorithm could infer the working state of a PL based on the correlation of measurements between neighbor VNs, which were mapped to both ends of the PL. Finally, the effectiveness and robustness of the proposed distributed online anomaly detection algorithms were verified on both the synthetic and real-world network datasets.

\appendices
\section{Proof of Proposition 1}
 As ${\bm{J}} = \bm{\Sigma} _{\bm{U}}^{ - 1/2} {\bm{R}}$ and $\bm{\Sigma} _{\bm{U}}^{ - 1/2}\bm{\Sigma} _{\bm{UY}} \bm{\Sigma} _{\bm{Y}}^{ - 1/2}  = {\bm{R}}\bm{\Sigma} {\bm{V}}^{\rm{T}}$, then ${\bm{R}}^{\rm{T}}\bm{\Sigma} _{\bm{U}}^{ - 1/2}=\bm{\Sigma}{\bm{V}}^{\rm{T}}\bm{\Sigma} _{\bm{Y}}^{ 1/2}\bm{\Sigma} _{\bm{UY}}^{-1} $ and ${\bm{J}}^{\rm{T}}{\bm{u}}=\bm{\Sigma}{\bm{V}}^{\rm{T}}\bm{\Sigma} _{\bm{Y}}^{ 1/2}\bm{\Sigma} _{\bm{UY}}^{-1} {\bm{u}}$. We can get that
\begin{equation}
\begin{gathered}
 {\bm{r}} = {\bm{J}}^{\rm{T}} {\bm{u}} - \bm{\Sigma}{\bm{L}}^{\rm{T}} {\bm{y}}
 \hfill \\
\;\;\;= \bm{\Sigma}{\bm{V}}^{\rm{T}} \bm{\Sigma} _{\bm{Y}}^{1/2}  \bm{\Sigma} _{\bm{UY}}^{ - 1} {\bm{u}} - \bm{\Sigma}{\bm{V}}^{\rm{T}} \bm{\Sigma} _Y^{ - 1/2} {\bm{y}}
  \hfill \\
 \;\;\; = \bm{\Sigma}{\bm{V}}^{\rm{T}} \bm{\Sigma} _{\bm{Y}}^{ - 1/2} (\bm{\Sigma} _{\bm{Y}} \bm{\Sigma} _{\bm{UY}}^{ - 1} {\bm{u}} - {\bm{y}}),
  \hfill \\
 \end{gathered}
\end{equation}
where ${\bm{\hat y}} = \bm{\Sigma} _{\bm{Y}} \bm{\Sigma} _{\bm{UY}}^{ - 1} {\bm{u}}$ is the least square estimation of $\bm{y}$ by $\bm{u}$. In this case, the residual has the minimal covariance value, and the $T_r^2$ statistic, which includes the inverse of the covariance matrix, has the best performance for anomaly detection \cite{jiang2017data}.
\section{Proof of Proposition 2}
The covariance calculation processes of $\bm{U}(t)$ are as the following steps.

Firstly, subtract average values from each column as
\begin{equation}
\begin{gathered}
\bar {\bm{U}}(t) = \left[ {\begin{array}{*{20}c}
   {u_{11}  - c_1 (t)} & {u_{12}  - c_2 (t)} &  \ldots  & {u_{1p}  - c_p (t)}  \\
   {u_{21}  - c_1 (t)} & {u_{22}  - c_2 (t)} &  \ldots  & {u_{2p}  - c_p (t)}  \\
    \vdots  &  \vdots  &  \ddots  &  \vdots   \\
   {u_{t1}  - c_1 (t)} & {u_{t2}  - c_2 (t)} &  \ldots  & {u_{tp}  - c_p (t)}  \\
\end{array}} \right]
 \hfill \\
 \;\;\;\;\;\;\;\;= \left[ {\begin{array}{*{20}c}
   {\bar u_{11} } & {\bar u_{12} } &  \ldots  & {\bar u_{1p} }  \\
   {\bar u_{21} } & {\bar u_{22} } &  \ldots  & {\bar u_{2p} }  \\
    \vdots  &  \vdots  &  \ddots  &  \vdots   \\
   {\bar u_{t1} } & {\bar u_{t2} } &  \ldots  & {\bar u_{tp} }
\end{array}} \right].
 \hfill \\
 \end{gathered}
\end{equation}

Then, the covariance of $\bm{U}(t)$  can be computed by
\begin{equation}
\begin{gathered}
 \bm{\Sigma} _{\bm{U}(t)}  = \frac{1}
{{t - 1}}(\bar {\bm{U}}(t))^{\rm{T}} \bar {\bm{U}}(t)  \hfill \\
 \;\;\;\;\;\;\;\;\;\;= \frac{1}{{t - 1}}\left[ {\begin{array}{*{20}c}
   {a_{11} } & {a_{12} } & {\rm{ \ldots }} & {a_{1p} }  \\
   {a_{21} } & {a_{22} } & {\rm{ \ldots }} & {a_{2p} }  \\
    \vdots  &  \vdots  &  \ddots  &  \vdots   \\
   {a_{p1} } & {a_{p2} } & {\rm{ \ldots }} & {a_{pp} }  \\
\end{array}} \right]. \\
 \end{gathered}
\end{equation}

The mean vector of $\bm{U}(t + 1)$ is $\left( {c_1 (t + 1),...,c_p (t + 1)} \right) = \left( {\frac{{tc_1 (t) + u_{(t + 1)1} }}{{t + 1}},...,\frac{{tc_3 (t) + u_{(t + 1)p} }}{{t + 1}}} \right)$, so
\begin{equation}
 \bar {\bm{U}}(t + 1) =
\left[ {\begin{array}{*{20}c}
   {\bar u_{11} {\rm{ + }}\frac{{c_1 (t) - u_{(t + 1)1} }}{{t + 1}}} &  \ldots  & {\bar u_{1p}  + \frac{{c_p (t) - u_{(t + 1)p} }}{{t + 1}}}  \\
   {\;\;\;\;\;\;\;\;\;\;\; \vdots } &  \ddots  & {\;\;\;\;\;\;\;\;\;\;\; \vdots }  \\
   {\frac{{t(u_{(t + 1)1}  - c_1 (t))}}{{t + 1}}} &  \ldots  & {\frac{{t(u_{(t + 1)p}  - c_p (t))}}{{t + 1}}}  \\
\end{array}} \right].
\end{equation}

Assume that the covariance matrix of $\bm{U}(t+1)$ is $\bm{\Sigma} _{\bm{U}(t+1)} = \frac{1}{t}\left[ {\begin{array}{*{20}c}
   {b_{11} } & {b_{12} } & {\rm{ \ldots }} & {b_{1p} }  \\
   {b_{21} } & {b_{22} } & {\rm{ \ldots }} & {b_{2p} }  \\
    \vdots  &  \vdots  &  \ddots  &  \vdots   \\
   {b_{p1} } & {b_{p2} } & {\rm{ \ldots }} & {b_{pp} }  \\
\end{array}} \right]$, where
\begin{equation}
\begin{gathered}
 b_{ij}  = \sum\limits_{k = 1}^t {\left( {\bar u_{ki} {\rm{ + }}\frac{{c_i (t) - u_{(t + 1)i} }}{{t + 1}}} \right)\left( {\bar u_{kj} {\rm{ + }}\frac{{c_j (t) - u_{(t + 1)j} }}{{t + 1}}} \right)}  \\
 \;\;\;\;\;\;\;\; + \left( {\frac{{t(u_{(t + 1)i}  - c_i (t))}}{{t + 1}}} \right)\left( {\frac{{t(u_{(t + 1)j}  - c_j (t))}}{{t + 1}}} \right) \\
 \;\;\; = a_{ij}  + \sum\limits_{k = 1}^t {\bar u_{ki} } \frac{{c_j (t) - u_{(t + 1)j} }}{{t + 1}} + \sum\limits_{k = 1}^t {\bar u_{kj} } \frac{{c_i (t) - u_{(t + 1)i} }}{{t + 1}} \\
 \;\;\;\;\;\; + t\left( {\frac{{c_i (t) - u_{(t + 1)i} }}{{t + 1}}} \right)\left( {\frac{{c_j (t) - u_{(t + 1)j} }}{{t + 1}}} \right) \hfill \\
 \;\;\;\;\;\; + \;\left( {\frac{{t(u_{(t + 1)i}  - c_i (t))}}{{t + 1}}} \right)\left( {\frac{{t(u_{(t + 1)j}  - c_j (t))}}{{t + 1}}} \right)\; \hfill \\
 \;\;\; = a_{ij}  + \;\frac{{t(c_i (t) - u_{(t + 1)i} )(c_j (t) - u_{(t + 1)j} )}}{{t + 1}} \hfill \\
 \;\;\;\;(\sum\limits_{k = 1}^t {\bar u_{ki} }  = 0,\sum\limits_{k = 1}^t {\bar u_{kj} }  = 0)(1 \le i,j \le p). \hfill \\
 \end{gathered}
\end{equation}

Therefore, the \emph{Proposition} 2 is proved.

\ifCLASSOPTIONcaptionsoff
  \newpage
\fi

\bibliographystyle{IEEEtran}
 \bibliography{IEEEabrv,bare_jrnl}

\end{document}